\documentclass[reprint,superscriptaddress,
showpacs,preprintnumbers,nofootinbib,
 amsmath,amssymb,aps,floatfix,prl
]{revtex4-1}

\usepackage{graphicx}
\usepackage{dcolumn}
\usepackage{bm}
\usepackage{hyperref}
\usepackage{cleveref}
\usepackage{dsfont}
\usepackage{placeins}
\usepackage{todonotes}
\usepackage[utf8]{inputenc}
\usepackage{gensymb}
\usepackage{multirow}
\usepackage{amsmath}
\usepackage{amsfonts}
\usepackage{amssymb}
\usepackage{array}   
\newcolumntype{L}{>{$}l<{$}} 
\newcolumntype{R}{>{$}r<{$}}
\newcolumntype{C}{>{$}c<{$}}
\usepackage{color}
\usepackage{xcolor}
\usepackage{xspace}
\usepackage{braket}
\usepackage{units}
\usepackage{slashed}
\usepackage[normalem]{ulem}
\usepackage[format=plain,justification=RaggedRight,singlelinecheck=false]{caption}
\usepackage[format=plain,justification=centering,singlelinecheck=false]{subcaption}

\DeclareMathOperator{\re}{Re}
\DeclareMathOperator{\im}{Im}

\newcommand{\eg}{{\it e.g.}\xspace}
\newcommand{\cf}{{\it cf.}\xspace}

\newcommand{\ie}{{\it i.e.}\xspace}

\newcommand{\etapi}{\ensuremath{\eta^{(\prime)}\pi}\xspace}

\newcommand{\mevnospace}{\ensuremath{{\mathrm{\,Me\kern -0.1em V}}}}
\newcommand{\gevnospace}{\ensuremath{{\mathrm{\,Ge\kern -0.1em V}}}}
\newcommand{\tevnospace}{\ensuremath{{\mathrm{\,Te\kern -0.1em V}}}}
\newcommand{\mev}{\mevnospace\xspace}
\newcommand{\gev}{\gevnospace\xspace}

\newcommand{\mevp}{\ensuremath{(\!\mevnospace)}}

\newcommand{\mytitle}[1]{\vspace{.5cm}{\em #1.---}}

\newcommand{\pione}
{\ensuremath{\pi_1}\xspace}

\newcommand{\pionemass}{\ensuremath{1564 \pm 24 \pm 86\mev}\xspace}
\newcommand{\pionewidth}{\ensuremath{492 \pm 54 \pm 102\mev}\xspace}

\allowdisplaybreaks

\hypersetup{ 
    pdfnewwindow=true,      
    colorlinks=true,       
    linkcolor=blue,          
    citecolor=blue,        
    filecolor=blue,      
    urlcolor=blue        
} 
\newcommand{\bonn}{Universit\"at Bonn, 
Helmholtz-Institut f\"ur Strahlen- und Kernphysik, 
53115 Bonn, Germany}
\newcommand{\ceem}{Center for  Exploration  of  Energy  and  Matter,  
Indiana  University,  
Bloomington,  IN  47403,  USA}
\newcommand{\ectstar}{European Centre for Theoretical Studies in Nuclear Physics and Related
Areas (ECT$^*$) and Fondazione Bruno Kessler,
I-38123 Villazzano (TN), Italy}
\newcommand{\ghent}{Department of Physics and Astronomy, Ghent University, Ghent 9000, Belgium}
\newcommand{\icn}{Instituto de Ciencias Nucleares, 
Universidad Nacional Aut\'onoma de M\'exico, 
Ciudad de M\'exico 04510, Mexico}
\newcommand{\indiana}{Physics  Department,  
Indiana  University,  
Bloomington,  IN  47405,  USA}
\newcommand{\jlab}{Theory Center,
Thomas  Jefferson  National  Accelerator  Facility, 
Newport  News,  VA  23606,  USA}
\newcommand{\mainz}{Institut f\"ur Kernphysik \& PRISMA Cluster of Excellence, 
Johannes Gutenberg Universit\"at, 
D-55099 Mainz, Germany}
\newcommand{\murcia}{Departamento de F\'isica, 
Universidad de Murcia, 
E-30071 Murcia, Spain}
\newcommand{\ucm}{Departamento de F\'isica Te\'orica, 
Universidad Complutense de Madrid, 
E-28040 Madrid, Spain}
\newcommand{\jpac}{Joint Physics Analysis Center}
\begin{document}
\title{Determination of the pole position of the lightest hybrid meson candidate}
\author{A.~Rodas}
\email{arodas@ucm.es}\affiliation{\ucm}
\author{A.~Pilloni}
\email{pillaus@jlab.org}\affiliation{\jlab}
\affiliation{\ectstar}
\author{M.~Albaladejo}
\affiliation{\jlab}
\affiliation{\murcia}
\author{C.~Fern\'andez-Ram\'irez}
\affiliation{\icn}
\author{A.~Jackura}
\affiliation{\ceem}
\affiliation{\indiana}
\author{V.~Mathieu}
\affiliation{\jlab}
\author{M.~Mikhasenko}
\affiliation{\bonn}
\author{J.~Nys}
\affiliation{\ghent}
\author{V.~Pauk}
\affiliation{\mainz}
\author{B.~Ketzer}
\affiliation{\bonn}
\author{A.~P.~Szczepaniak}
\affiliation{\jlab}\affiliation{\ceem}\affiliation{\indiana}
\collaboration{\jpac}
\noaffiliation
\preprint{JLAB-THY-18-2839}
\begin{abstract}
Mapping states with explicit gluonic degrees of freedom in the light sector is a challenge,
and has led to controversies in the past. In particular, the experiments have reported two different hybrid candidates
with spin-exotic signature, $\pi_1(1400)$ and $\pi_1(1600)$, which couple separately 
to $\eta \pi$ and $\eta' \pi$. This picture is not compatible with recent Lattice QCD estimates for hybrid states, nor with most phenomenological models.
We consider the recent partial wave analysis of the \etapi system by the COMPASS collaboration.
We fit the extracted intensities and phases with a coupled-channel amplitude that enforces the unitarity and analyticity of the $S$-matrix. 
We provide
a robust extraction of a single exotic \pione 
resonant pole, with mass and width \pionemass and \pionewidth, which couples to both \etapi channels. We find no evidence for a second exotic state. 
We also provide the resonance parameters of the $a_2(1320)$ and $a'_2(1700)$.
\end{abstract}

\maketitle

\mytitle{Introduction} 
Explaining the structure of hadrons in terms of quarks and gluons,  
the fundamental building blocks of Quantum Chromodynamics (QCD), is of key importance to our understanding of strong interactions.
The vast majority of observed mesons can be classified as 
$q\bar q$ bound states, although QCD 
should have, in principle, a much richer spectrum. 
Indeed, 
several experiments have reported resonance candidates 
that do not fit the valence quark model template~\cite{Ketzer:2012vn,Meyer:2015eta}, mainly in the heavy sector~\cite{Esposito:2016noz,Lebed:2016hpi,Guo:2017jvc,Olsen:2017bmm,Karliner:2017qhf}. 
These new experimental results, together with rapid advances in lattice 
gauge computations,  open new fronts in studies of the fundamental aspects of QCD, such as quark confinement and 
mass generation.  Since gluons are the mediators of  the strong interaction, QCD dynamics cannot be fully understood without addressing the role of gluons in binding hadrons. 
The existence of states with explicit excitations of the gluon field, commonly referred to as {\em hybrids}, 
was postulated a long time ago~\cite{Horn:1977rq,Isgur:1984bm,Chanowitz:1982qj,Barnes:1982tx,Close:1994hc}, and has
 recently been supported by lattice~\cite{Lacock:1996ny,Bernard:1997ib,Dudek:2013yja}
and phenomenological QCD studies~\cite{Szczepaniak:2001rg,Szczepaniak:2006nx,Guo:2008yz,Bass:2018xmz}. 
In particular, a state with exotic quantum numbers $J^{PC} (I^G) = 1^{-+} (1^-)$ in the mass range $1.7 - 1.9\gev$ is generally expected. 
The 
experimental determination of hybrid hadron
properties ---\eg their masses, widths, and decay patterns--- provides a unique opportunity for a systematic study of low-energy gluon dynamics.
This has motivated the COMPASS spectroscopy program~\cite{Baum:1996yv,Abbon:2014aex} and 
the 12\gev upgrade of Jefferson Lab, with experiments dedicated to hybrid photoproduction at CLAS12 and GlueX~\cite{Rizzo:2016qvl,Dobbs:2017vjw}.

The hunt for hybrid mesons
is challenging, since the 
spectrum of particles produced in high energy collisions 
 is dominated by nonexotic resonances. 
 The extraction of exotic signatures 
 requires sophisticated partial-wave amplitude analyses. In the past, inadequate model assumptions and limited statistics resulted in debatable results.
The first reported hybrid candidate was the $\pi_1(1400)$ in the $\eta \pi$ final state~\cite{Thompson:1997bs,Chung:1999we,Adams:2006sa,Abele:1998gn,Abele:1999tf}. Another state, the $\pi_1(1600)$,
was later claimed in the $\rho \pi$ and $\eta'\pi$ channels, with different resonance 
parameters~\cite{Ivanov:2001rv,Khokhlov:2000tk}. 
The COMPASS experiment 
confirmed a 
peak in $\rho\pi$ and $\eta'\pi$ at around $1.6\gev$~\cite{Alekseev:2009aa,Akhunzyanov:2018pnr} 
and an additional structure in $\eta \pi$, at approximately   $1.4\gev$~\cite{Adolph:2014rpp}.
A theoretical approach based on a unitarized $U(1)$-extended chiral Lagrangian predicted a $\pi_1$ state with mass of about $1400\mev$ decaying mostly into $\eta'\pi$~\cite{Bass:2001zs}.
A phenomenological unitary coupled-channel 
analysis of the $\etapi$ system from E852 data was instead not conclusive~\cite{Szczepaniak:2003vg}.
While the $\pi_1(1600)$ is close to the expectation for a hybrid, the observation of two nearby $1^{-+}$ hybrids below \mbox{2\gev} is surprising.  This makes 
 the microscopic interpretation of the $\pi_1(1400)$ problematic.   
Moreover, in the $SU(3)$ limit, Bose symmetry  prevents the decay of a hybrid into $\eta \pi$~\cite{Close:1987aw}. A tetraquark interpretation of the lighter state  might be viable, and would explain why this state has eluded predictions in constituent gluon models. However, this interpretation would lead to the prediction of unobserved doubly charged and doubly strange mesons~\cite{Chung:2002fz}, and is unfavored in the diquark-antidiquark model~\cite{Jaffe:2003sg,Jaffe:2004ph}.
Establishing whether there exists one or two exotic states in this mass region 
is thus a stringent test for the available phenomenological frameworks in the nonperturbative regime.

In~\cite{Jackura:2017amb} we analyzed the spectrum of the  $\eta \pi$ $D$-wave extracted from the COMPASS data. 
In this Letter, we extend the mass dependent study to the exotic $P$-wave, and present results of the first coupled-channel analysis of the $\etapi$ COMPASS data. 
We establish that a single exotic \pione is needed and provide a detailed analysis of its properties.
We also determine the resonance parameters of the nonexotic $a_2(1320)$ and $a_2'(1700)$.

\begin{figure*}
\centering
\includegraphics[width=.32\textwidth]{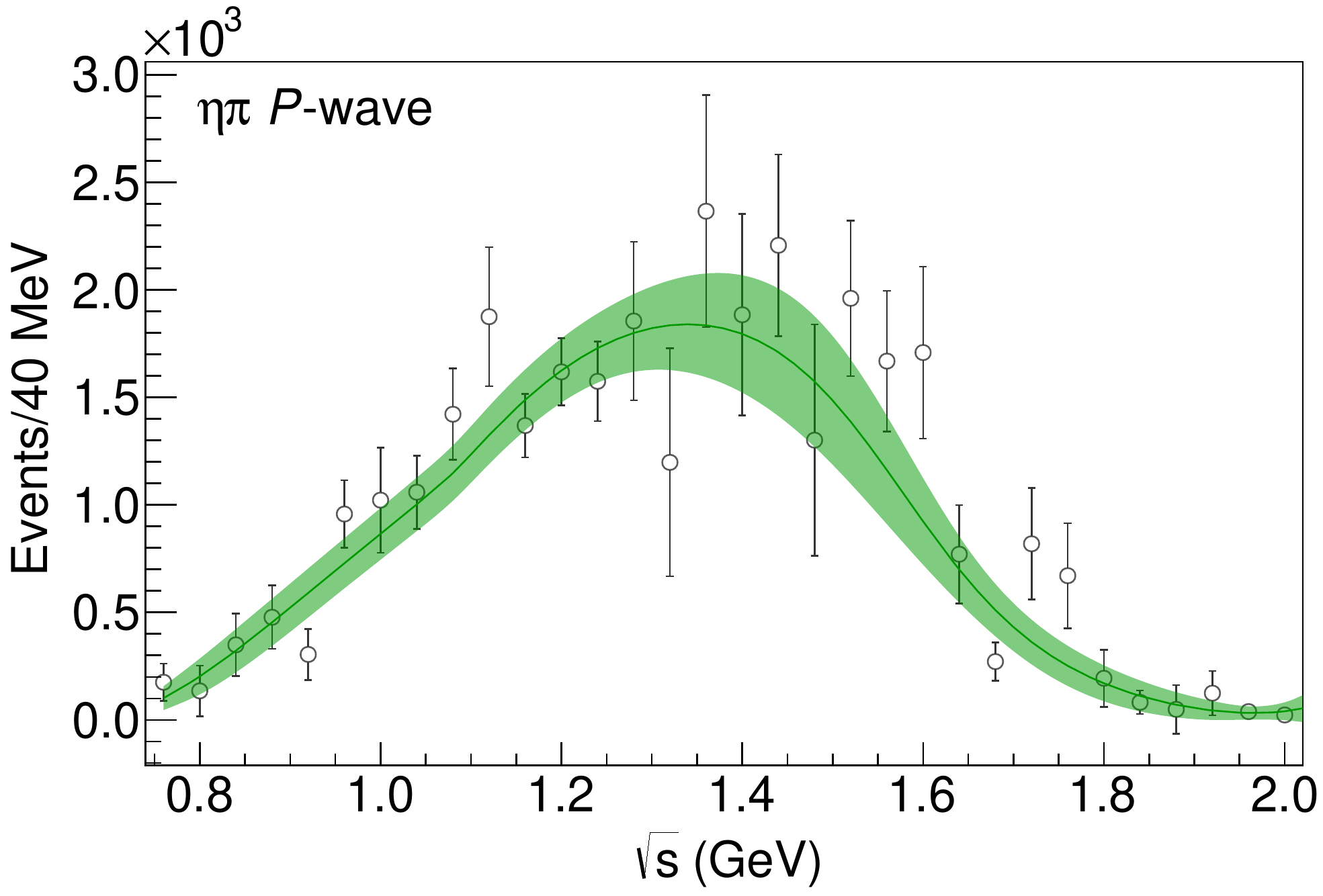}
\includegraphics[width=.32\textwidth]{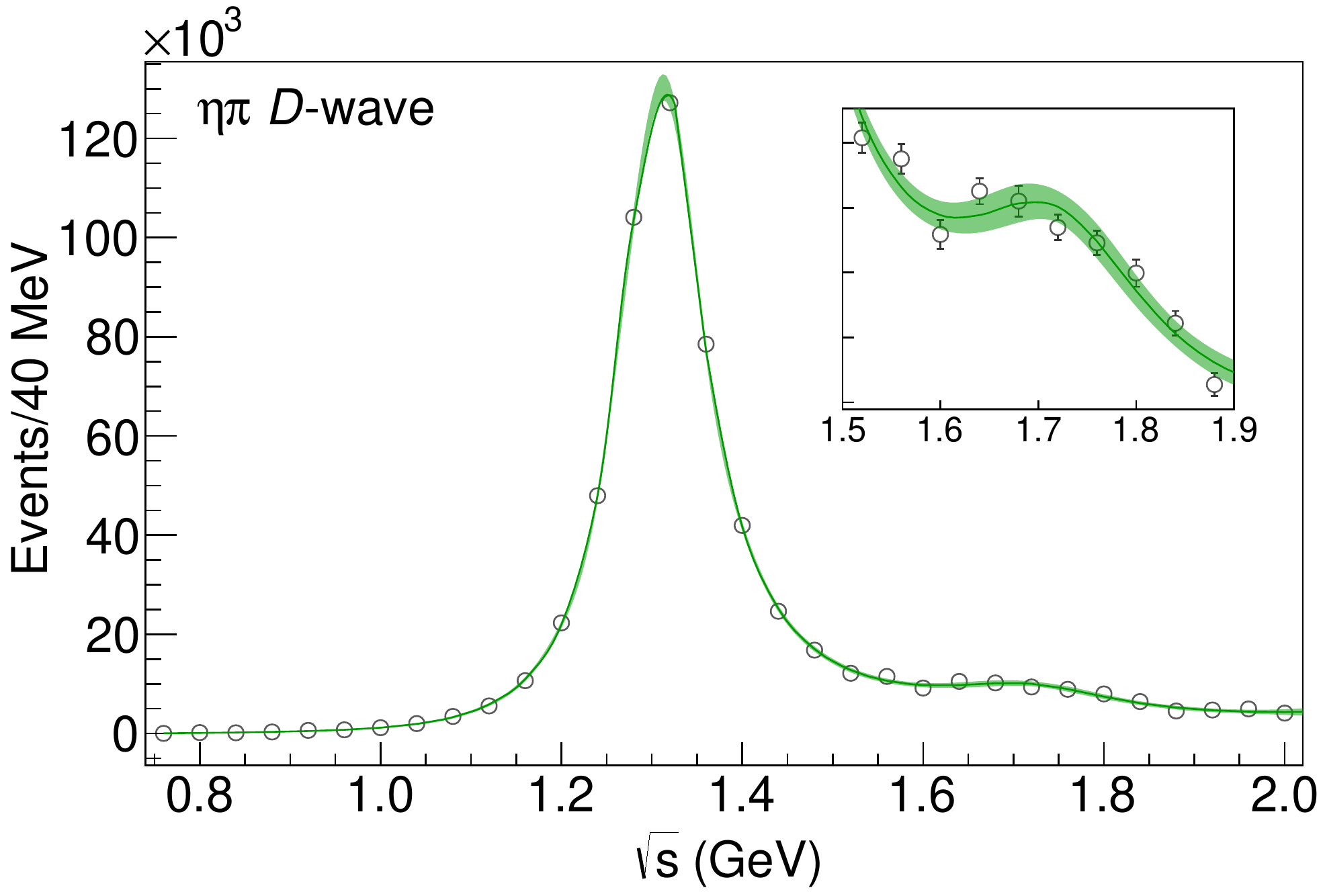}
\includegraphics[width=.32\textwidth]{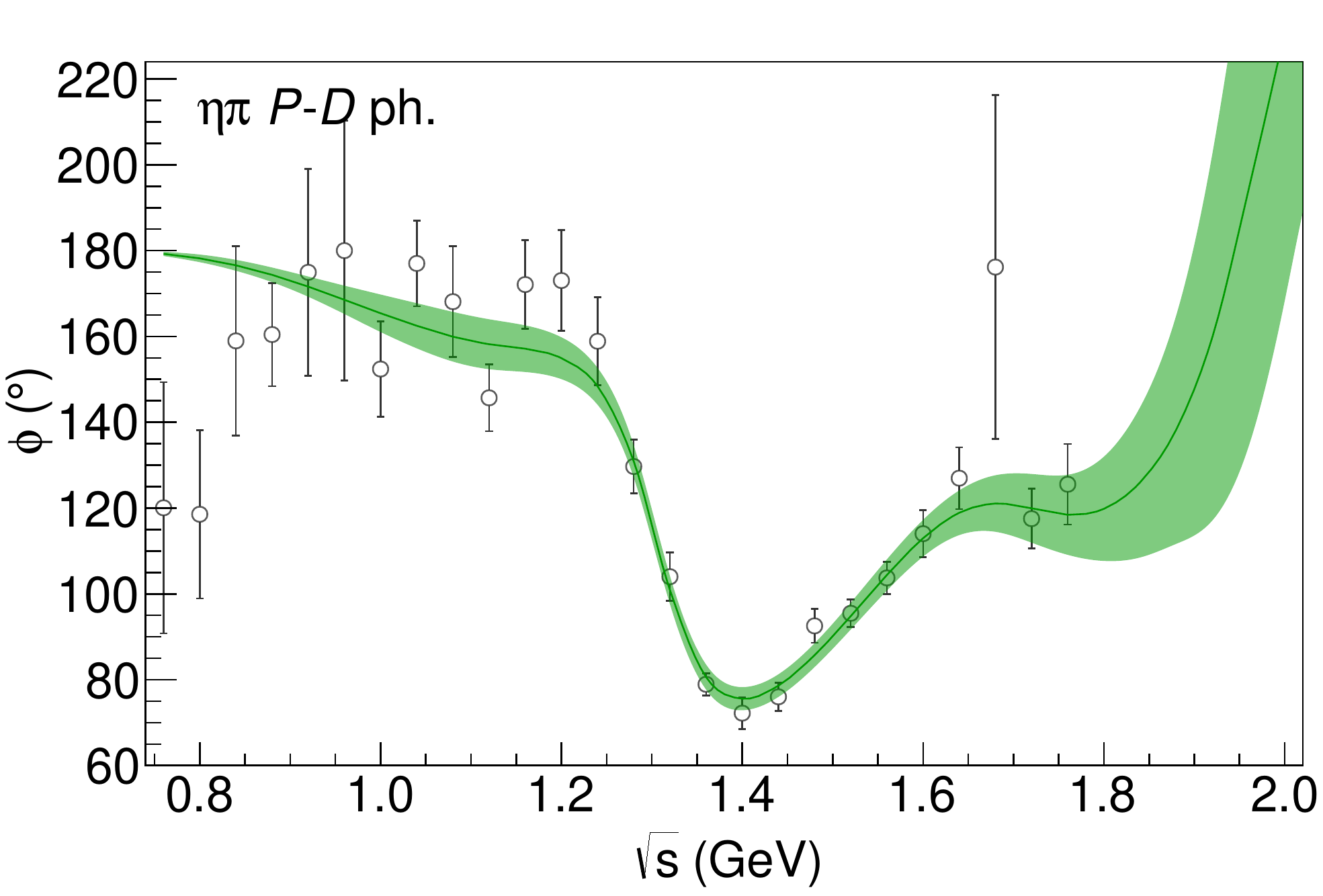}\\
\includegraphics[width=.32\textwidth]{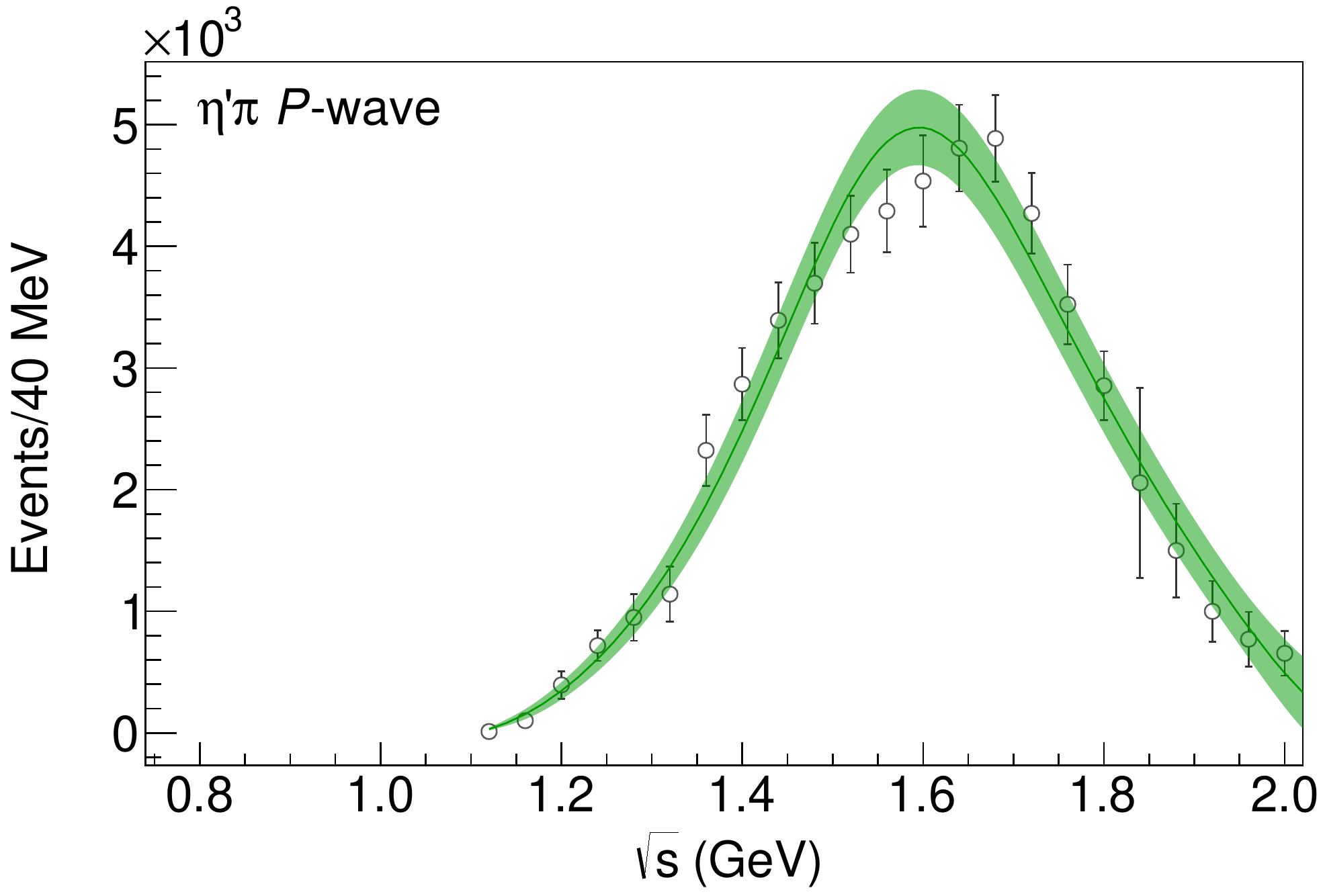}
\includegraphics[width=.32\textwidth]{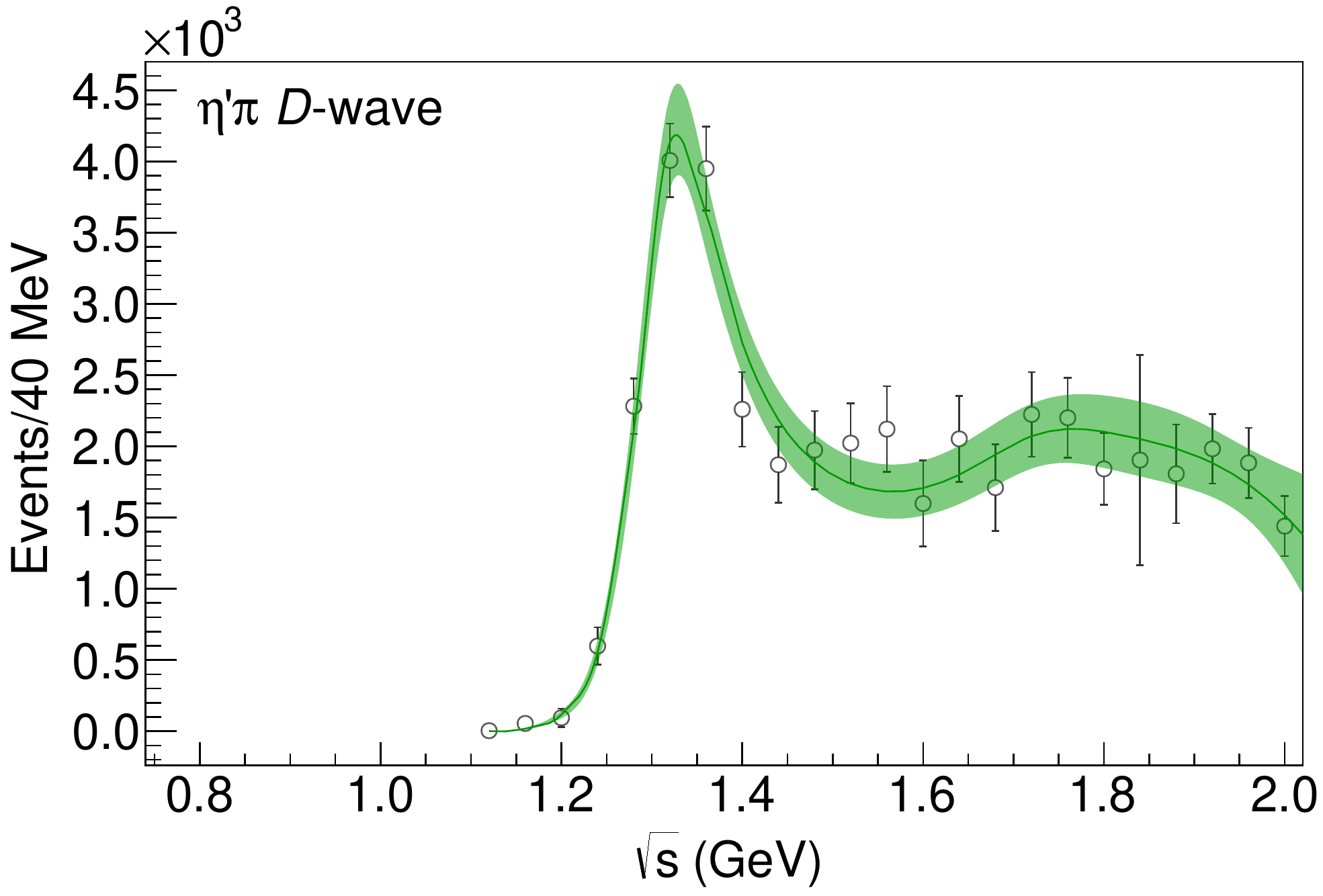}
\includegraphics[width=.32\textwidth]{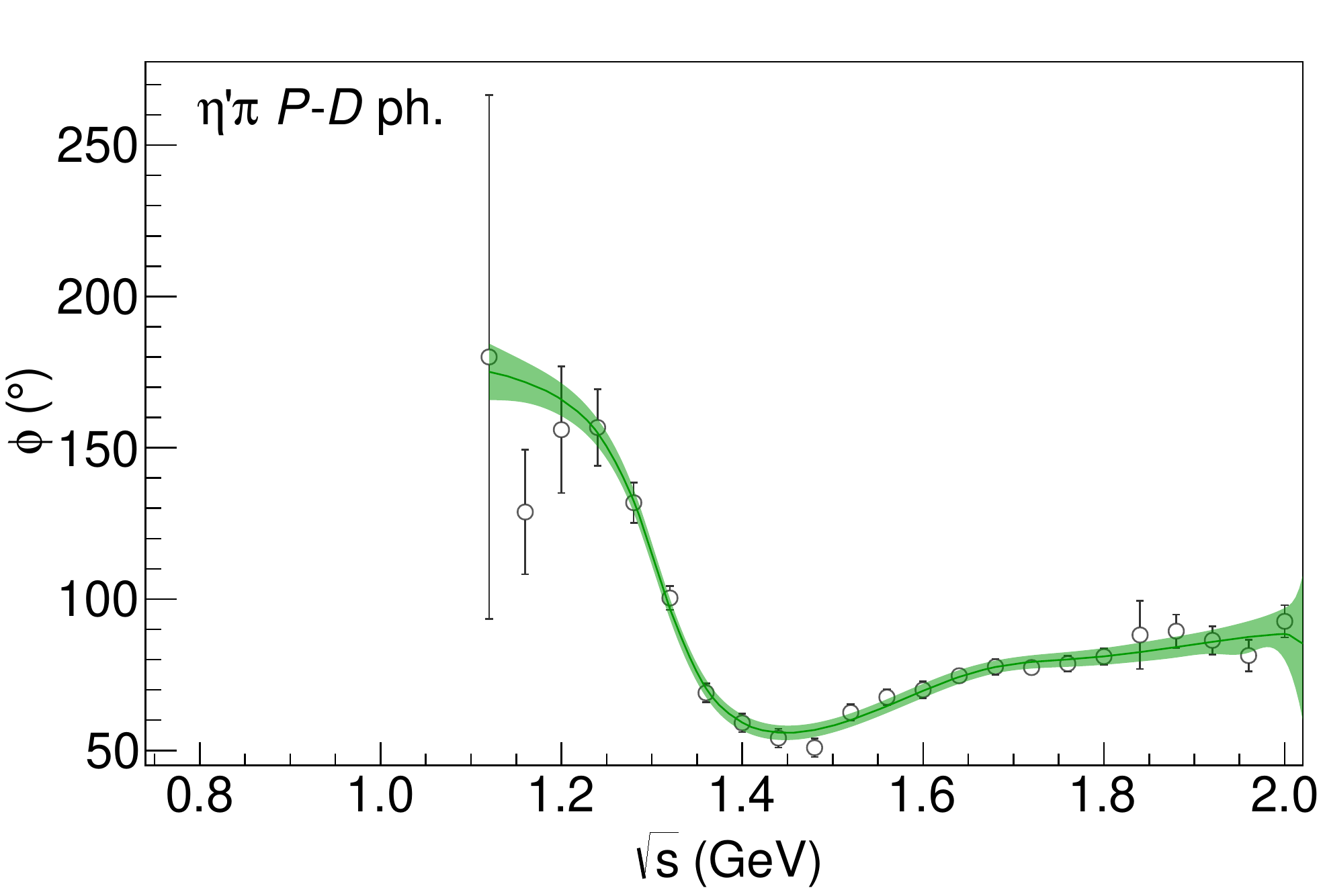}
\caption{Fits to the $\eta \pi$ (upper line) and $\eta' \pi$ (lower line) data from COMPASS~\cite{Adolph:2014rpp}. 
The intensities of $P$- (left), $D$-wave (center), and their relative phase
(right) are shown. The inset zooms into the region of the $a_2'(1700)$. The solid line and green band show the result of the fit and the $2\sigma$ confidence level provided by the bootstrap analysis, respectively. The initialization of the fit is chosen by randomly generating $O(10^5)$ different sets of values for the parameters. The best fit has $\chi^2/\text{dof} = 162/122=1.3$. The errors shown are statistical only.
 } \label{fig:fits}
\end{figure*}

\mytitle{Description of the data}
We use the mass independent analysis by COMPASS of 
$\pi p \to \etapi p$, with a $190\gev$ 
pion beam~\cite{Adolph:2014rpp}.
We focus on the $P$- and $D$-wave intensities and their relative phase, in both channels. 
The published data are integrated over the range of transferred momentum squared  
$-t_1 \in [0.1, 1]\gev^2$.
However, given the diffractive nature of the reaction, most of the events are produced in the forward direction, near the lower limit in $-t_1$.  
 The $\etapi$ partial-wave intensities and phase differences are given in 40\mev mass bins, from threshold up to 3\gev. Intensities are normalized 
to the number of observed events corrected by 
the detector acceptance. 
The errors quoted are 
statistical only; systematic uncertainties or correlations in the extraction of the partial waves were found to be negligible~\cite{Schluter:2012mep}. 
We thus assume that all data points are 
independent and normally distributed. 
As seen in Figs. 4(a) and 5(a) of~\cite{Adolph:2014rpp},  at the  $\eta' \pi$  mass of $2.04\gev$ there is a sharp falloff  
 in the $P$-wave intensity,  
  and a sudden change by  $50\degree$ in the phase difference between the $P$- and $D$-wave.  
  This behavior might be attributed to another state. The E852 experiment claimed indeed a third exotic $\pi_1(2015)$ in the $f_1(1285)\pi$ and $\omega\pi\pi$ channels~\cite{Kuhn:2004en,Lu:2004yn}. However, this state 
   is  too broad to explain such an abrupt behavior  and thus it is difficult to find a compelling explanation. 
 Unfortunately it is not possible to crosscheck this behavior with the $\eta \pi$ relative phase due to lack of data in the $1.8-2.0\gev$ region.
Moreover, fitting these features of the $P$-wave 
drives the position of the $a_2'$ to unphysical values. For these 
reasons, we fit data up to $2\gev$ only.

Enforcing unitarity allows us to properly implement the interference among the various resonances and the background. In principle, one wishes to include all the kinematically allowed channels in a unitary analysis.
Recently, COMPASS  published the complete $3\pi$ partial-wave analysis~\cite{Akhunzyanov:2018pnr}, including the exotic $1^{-+}$ wave in the $\rho\pi$ final state. 
However, the extraction of the resonance pole in this channel is hindered by the irreducible Deck process~\cite{Deck:1964hm}, which refers to the exchange of a pion between the final state $\rho$ and $\pi$ (\cf Figure 8 in~\cite{Akhunzyanov:2018pnr}). This generates a peaking background in the exotic partial wave~\cite{Ascoli:1974hi,dimadeck}. 
Since the Deck mechanism is not fully accounted for in the COMPASS amplitude model, we  do not include the $3\pi$ data in our analysis. 
 As discussed in~\cite{Jackura:2017amb}, neglecting additional 
channels 
 does not affect the pole position, 
as long as the resonance poles are far away from threshold, which 
 is the case studied here. 

\mytitle{Reaction model}
At high energies, peripheral production of $\pi p\to \etapi p$ is dominated by Pomeron~($\mathbb{P}$) exchange. 
The notion of Pomeron exchange emerges from Regge theory~\cite{Chew:1961ev,Donnachie:2002en}, and allows us to factorize the $\pi\mathbb{P}\to \etapi$ process. 
For fixed $t_1$ the latter resembles an ordinary helicity partial wave amplitude  $a^J_{i}(s)$, with $i=\etapi$ the final channel index, $J$  the angular momentum of the  \etapi system and $s$ its invariant mass squared. This amplitude, in principle, 
 also depends on the effective spin and helicity of the Pomeron. 
However, the approximately constant hadron cross section at high energies implies that the effective spin 
 of the Pomeron is near one, which 
explains dominance of the partial wave components with angular momentum projection $M=\pm 1$ as seen in data~\cite{Close:1999is,Arens:1996xw,Adolph:2014rpp}. Since the two are related by parity, we drop reference to the Pomeron quantum numbers (for more details, see~\cite{Jackura:2017amb}).
As discussed previously, we fix an effective value $t_\text{eff}=-0.1\gev^2$.

We parameterize the 
 amplitudes following the coupled-channel $N/D$ 
formalism~\cite{Bjorken:1960zz},
\begin{equation}
\label{eq:amplitude}
 a^J_i(s) = q^{J-1} p_i^J \, \sum_k n^J_k(s) \left[ {D^J(s)}^{-1} \right]_{ki}\,,
\end{equation}
where  
$p_i$ 
is the \etapi breakup momentum, and  
$q$ 
the $\pi$ beam momentum in the \etapi rest frame.
\footnote{One unit of incoming momentum $q$ is divided out because of the Pomeron-nucleon vertex~\cite{Jackura:2017amb}.}
The $n^J_k(s)$'s incorporate   exchange ``forces" in the production process   
and are smooth functions of $s$ in the physical region. 
The $D^J(s)$ matrix represents the $\etapi \to \etapi$ final state interactions, and contains cuts  on the real axis 
above thresholds  (right hand cuts), which are  constrained by  unitarity.

For the numerator $n^J_k(s)$, we use an effective expansion in Chebyshev polynomials. 
A customary parameterization of the denominator is given by~\cite{Aitchison:1972ay}
\begin{equation}\label{eq:Dsol}
D^J_{ki}(s) =  \left[ {K^J(s)}^{-1}\right]_{ki} - \frac{s}{\pi}\int_{s_{k}}^{\infty}ds'\frac{\rho N^J_{ki}(s') }{s'(s'-s - i\epsilon)}, 
\end{equation}
where 
$s_k$ is the threshold in channel $k$ and
\begin{align}
\rho N^J_{ki}(s') &= \delta_{ki} \,\frac{\lambda^{J+1/2}\left(s',m_{\eta^{(\prime)}}^2,m_\pi^2\right)}{\left(s'+s_L\right)^{2J+1+\alpha}} \label{eq:rhoN}
\intertext{is an effective description of the left 
hand singularities in the $\etapi\to\etapi$ scattering, which is 
controlled by the $s_L$ parameter fixed at the hadronic scale  $\simeq 1\gev^2$. Finally, 
}
K^J_{ki}(s) &= \sum_R \frac{g^{J,R}_k g^{J,R}_i}{m_R^2 - s} + c^J_{ki} + d^J_{ki} \,s,\label{eq:Kmatrix}
\end{align}
with $c^J_{ki}= c^J_{ik}$ and $d^J_{ki}= d^J_{ik}$,
is a standard parameterization for the $K$-matrix.
In our reference model, we consider two $K$-matrix poles in the $D$-wave, and one single $K$-matrix pole in the $P$-wave; 
the numerator of each channel and wave is described by a third-order polynomial.
We set $\alpha = 2$ in Eq.~\eqref{eq:rhoN}, which has been effective in describing the single-channel case~\cite{Jackura:2017amb}. The remaining 37 parameters are fitted to data, by performing a $\chi^2$ minimization with {\tt MINUIT}~\cite{minuit}. 
As shown in Fig.~\ref{fig:fits}, the result of the best fit is in good agreement with data. In particular, a single $K$-matrix pole is able to correctly describe the $P$-wave peaks in the two channels,  which are separated by 
  $200\mev$.
The shift of the peak in the 
 $\eta\pi$ spectrum to lower energies 
  originates from the combination  between final state interactions and the production process. 
The uncertainties on the parameters are estimated via the bootstrap method~\cite{recipes,EfroTibs93}:  we generate a large number of pseudo datasets and refit each one of them. The (co)variance of the parameters provides an estimate of their statistical uncertainties and correlations. The values of the fitted parameters and their covariance matrix are provided in the Supplemental Material~\cite{suppl}. The average curve passes the Gaussian test in~\cite{Perez:2015pea}.

\begin{figure*}
\includegraphics[width=\textwidth]{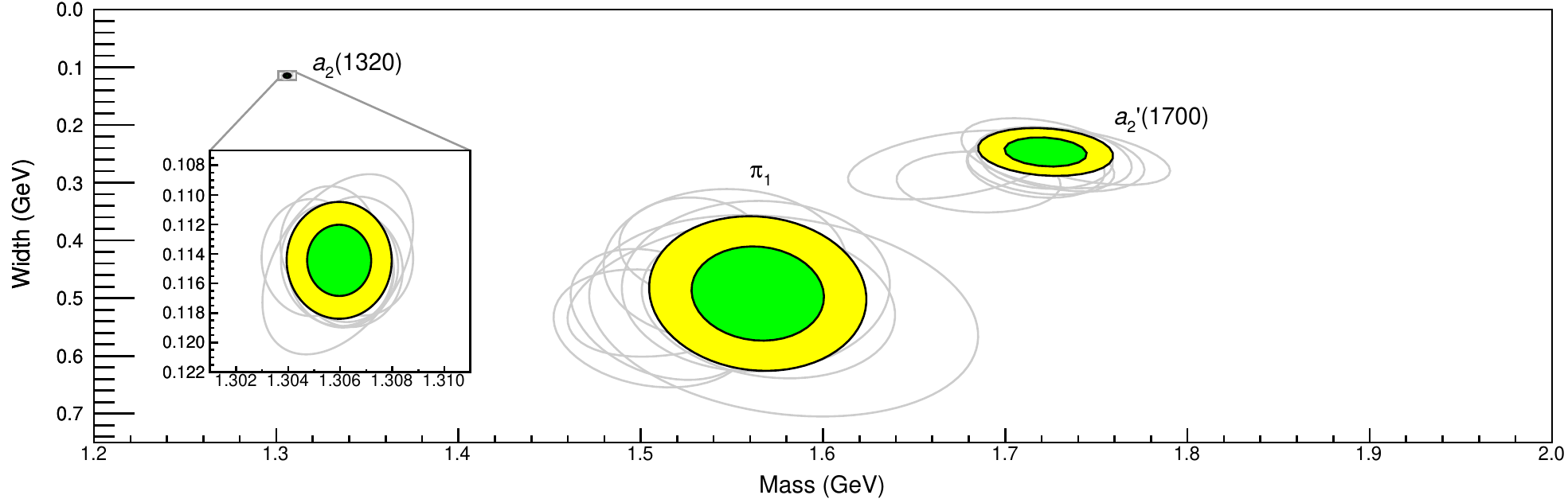}
\caption{\label{fig:poles} Positions of the poles identified as the $a_2(1320)$, \pione, and $a_2'(1700)$. The inset shows the position of the $a_2(1320)$. The green and yellow ellipses show the $1\sigma$ and $2\sigma$ confidence levels, respectively. The gray ellipses in the background show, 
 within $2\sigma$, variation of the pole position  upon changing the functional form and the parameters of the model, as discussed in the text
 }
\end{figure*}
Once the parameters are determined, the amplitudes can be analytically continued to complex values of $s$.
The $D^J(s)$ matrix in Eq.~\eqref{eq:Dsol} can be continued underneath 
the unitarity cut into the closest unphysical Riemann sheet. A pole $s_P$ in the amplitude appears when the determinant of $D^J(s_P)$ vanishes. Poles close to the real axis influence the physical region and can be identified as resonances, whereas further singularities are likely to be artifacts of the specific model with no direct physical interpretation. 
For any practical parameterization, especially in a coupled-channel problem, it is not possible to  specify {\it a priori} the number of poles. Appearance of spurious poles far from the physical region is thus unavoidable. It is however possible to isolate the physical poles by testing their stability against different parameterizations and data resampling. 
We select the resonance poles in the $m\in [1,2]\gev$ and $\Gamma \in [0,1]\gev$ region, where customarily $m = \re \sqrt{s_P}$ and $\Gamma = -2\im \sqrt{s_P}$.
We find two poles in the $D$-wave, identified as the $a_2(1320)$ and $a_2'(1700)$, and a single pole in the $P$-wave, which we call \pione. The pole positions are shown in Fig.~\ref{fig:poles}, and the resonance parameters in Table~\ref{tab:poles}. 
To estimate the statistical significance of the $\pi_1$ pole, we perform fits using a pure background model for the $P$-wave, \ie setting $g^{P,1}_{\etapi} = 0$ in Eq.~\eqref{eq:Kmatrix}. 
The best solution having no poles in our reference region has a 
$\chi^2$ almost 50 times larger, which rejects the possibility for the $P$-wave peaks to be generated by nonresonant production.
We also considered solutions having two isolated $P$-wave  poles in the reference region, which would correspond to the scenario discussed 
 in the PDG~\cite{pdg}. The $\chi^2$ for this case is equivalent to the single pole solution. One of the poles is compatible with the previous  determination, while the second is unstable, \ie it can appear in a large region of the $s$-plane depending on the initial values of the fit parameters. 
 Moreover, the behavior of the $\eta \pi$ phase required by the fit is rather peculiar. A $180^\circ$ jump (due to a zero in the amplitude) appears  above $1.8 \gev$, where no data exist. We conclude there is no evidence for a second pole.

\mytitle{Systematic uncertainties}
Unlike the COMPASS mass independent fit, the pole extraction carries systematic uncertainties associated with the reaction model. To assess these, 
we vary the parameters and functional forms which were kept fixed in the previous fits.  We can separate these in two categories: $i)$ variations  of the  numerator function $n^J_k(s)$ in Eq.~\eqref{eq:amplitude},  which is expected to be
 smooth in the region of the data,  and $ii)$ variations of the function $\rho N(s')$, which determines the imaginary part of the denominator in Eq.~\eqref{eq:Dsol}. 
 As for the latter, we investigate whether the specific form we chose biases the determination of the poles.
 Upon variation of the parameters and of the functional forms, the shape of the dispersive integral in Eq.~\eqref{eq:Dsol} is altered, but the fit quality is unaffected. The pole positions change roughly within $2\sigma$, as one can see in Fig.~\ref{fig:poles}.   
As for the numerator $n^J_k(s)$, we varied the effective value of $t_\text{eff}$ and  the order of the polynomial expansion. 
Given the flexibility of the numerator parameterization, these variations effectively absorb the model dependence related to the production mechanism. None of these cause important changes in 
 pole locations. 
   Our final estimate for the uncertainties is reported in Table~\ref{tab:poles}, while the detailed summary is given in the Supplemental Material~\cite{suppl}.

\begin{table}[t] 
\caption{Resonance parameters. The first error is statistical, the second systematic.}
\begin{ruledtabular}
\begin{tabular}{c c c} 
Poles & Mass \mevp & Width \mevp \\ \hline 
$a_2(1320)$ & $1306.0 \pm 0.8 \pm 1.3 $ & $114.4 \pm 1.6 \pm 0.0$ \\ 
$a_2'(1700)$ & $1722 \pm 15 \pm 67 $ & $247 \pm 17 \pm 63$ \\ 
$\pione$ & $1564 \pm 24 \pm 86 $ & $492 \pm 54 \pm 102$ \\ 
\end{tabular}

\label{tab:poles}
\end{ruledtabular}
\end{table}

\mytitle{Conclusions} We performed the first coupled-channel analysis of the  $P$- and $D$-waves in the $\etapi$ system measured at COMPASS~\cite{Adolph:2014rpp}. We used an amplitude parameterization constrained by unitarity and analyticity. 
We find  two poles in the $D$-wave, which we identify as the $a_2(1320)$ and the $a_2'(1700)$, with resonance parameters  consistent with the single-channel analysis~\cite{Jackura:2017amb}. 
In the $P$-wave, we find a single exotic \pione in the region constrained by  data.  
 This determination  is compatible with the existence of a single isovector hybrid meson with quantum numbers $J^{PC}=1^{-+}$, as suggested by lattice QCD~\cite{Lacock:1996ny,Bernard:1997ib,Dudek:2013yja}. Its mass and width are determined to be \pionemass and \pionewidth, respectively.  
The statistical uncertainties are estimated via the bootstrap technique, while the systematics due to model dependence are assessed by varying  parameters and  functional forms that are not directly constrained by unitarity. 
We find no evidence for a  second pole that could be identified 
 with another  $\pi_1$ resonance. Solutions with two poles are possible, but do not improve the fit quality and, when present, 
   the position of the second pole 
  is  unstable against different starting values of the fit. 
It is worth noting that the two-pole solutions have a 
  peculiar behavior of the $\eta\pi$ phase in the $\gtrsim 2\gev$ mass region, where no data exist. 
 New data from GlueX and CLAS12 experiments at Jefferson Lab in this and  higher mass region will be valuable to test this behavior.

\begin{acknowledgments} \mytitle{Acknowledgments}
We would like to thank the COMPASS collaboration for useful comments. AP thanks the Mainz Institute for Theoretical Physics (MITP) for its kind hospitality while this work was being completed.
This work was supported by
the U.S.~Department of Energy under Grants
No.~DE-AC05-06OR23177 
and No.~DE-FG02-87ER40365, 
U.S.~National Science Foundation under Grant No.
PHY-1415459, 
Ministerio de Ciencia, Innovaci\'on y Universidades (Spain)
Grants No.~FPA2016-75654-C2-2-P and No.~FPA2016-77313-P, Universidad Complutense de Madrid predoctoral scholarship program, 
Research Foundation -- Flanders (FWO), 
PAPIIT-DGAPA (UNAM, Mexico) under Grants No.~IA101717 and
No.~IA101819, 
CONACYT (Mexico) under grant No.~251817, 
the German Bundesministerium f\"{u}r Bildung und Forschung (BMBF), 
and Deutsche Forschungsgemeinschaft (DFG)
through the Collaborative Research Center 
[The Low-Energy Frontier of the Standard Model (SFB 1044)] 
and the Cluster of Excellence 
[Precision Physics, Fundamental Interactions and Structure of Matter (PRISMA)].
\end{acknowledgments}
\bibliographystyle{apsrev4-1-jpac}
\bibliography{quattro.bib}

\begin{thebibliography}{58}%
\makeatletter
\providecommand \@ifxundefined [1]{%
 \@ifx{#1\undefined}
}%
\providecommand \@ifnum [1]{%
 \ifnum #1\expandafter \@firstoftwo
 \else \expandafter \@secondoftwo
 \fi
}%
\providecommand \@ifx [1]{%
 \ifx #1\expandafter \@firstoftwo
 \else \expandafter \@secondoftwo
 \fi
}%
\providecommand \natexlab [1]{#1}%
\providecommand \enquote  [1]{``#1''}%
\providecommand \bibnamefont  [1]{#1}%
\providecommand \bibfnamefont [1]{#1}%
\providecommand \citenamefont [1]{#1}%
\providecommand \href@noop [0]{\@secondoftwo}%
\providecommand \href [0]{\begingroup \@sanitize@url \@href}%
\providecommand \@href[1]{\@@startlink{#1}\@@href}%
\providecommand \@@href[1]{\endgroup#1\@@endlink}%
\providecommand \@sanitize@url [0]{\catcode `\\12\catcode `\$12\catcode
  `\&12\catcode `\#12\catcode `\^12\catcode `\_12\catcode `\%12\relax}%
\providecommand \@@startlink[1]{}%
\providecommand \@@endlink[0]{}%
\providecommand \url  [0]{\begingroup\@sanitize@url \@url }%
\providecommand \@url [1]{\endgroup\@href {#1}{\urlprefix }}%
\providecommand \urlprefix  [0]{URL }%
\providecommand \Eprint [0]{\href }%
\providecommand \doibase [0]{http://dx.doi.org/}%
\providecommand \selectlanguage [0]{\@gobble}%
\providecommand \bibinfo  [0]{\@secondoftwo}%
\providecommand \bibfield  [0]{\@secondoftwo}%
\providecommand \translation [1]{[#1]}%
\providecommand \BibitemOpen [0]{}%
\providecommand \bibitemStop [0]{}%
\providecommand \bibitemNoStop [0]{.\EOS\space}%
\providecommand \EOS [0]{\spacefactor3000\relax}%
\providecommand \BibitemShut  [1]{\csname bibitem#1\endcsname}%
\let\auto@bib@innerbib\@empty
\bibitem [{\citenamefont {Ketzer}(2012)}]{Ketzer:2012vn}%
  \BibitemOpen
  \bibfield  {author} {\bibinfo {author} {\bibfnamefont {B.}~\bibnamefont
  {Ketzer}},\ }\href {\doibase 10.22323/1.157.0025} {\bibfield  {journal}
  {\bibinfo  {journal} {PoS}\ }\textbf {\bibinfo {volume} {QNP2012}},\ \bibinfo
  {pages} {025} (\bibinfo {year} {2012})},\ \Eprint
  {http://arxiv.org/abs/1208.5125}{\tt arXiv:1208.5125 [hep-ex]}\BibitemShut
  {NoStop}%
\bibitem [{\citenamefont {Meyer}\ and\ \citenamefont
  {Swanson}(2015)}]{Meyer:2015eta}%
  \BibitemOpen
  \bibfield  {author} {\bibinfo {author} {\bibfnamefont {C.~A.}\ \bibnamefont
  {Meyer}}\ and\ \bibinfo {author} {\bibfnamefont {E.~S.}\ \bibnamefont
  {Swanson}},\ }\href {\doibase 10.1016/j.ppnp.2015.03.001} {\bibfield
  {journal} {\bibinfo  {journal} {Prog.Part.Nucl.Phys.}\ }\textbf {\bibinfo
  {volume} {82}},\ \bibinfo {pages} {21} (\bibinfo {year} {2015})},\ \Eprint
  {http://arxiv.org/abs/1502.07276}{\tt arXiv:1502.07276 [hep-ph]}\BibitemShut
  {NoStop}%
\bibitem [{\citenamefont {Esposito}\ \emph {et~al.}(2016)\citenamefont
  {Esposito}, \citenamefont {Pilloni},\ and\ \citenamefont
  {Polosa}}]{Esposito:2016noz}%
  \BibitemOpen
  \bibfield  {author} {\bibinfo {author} {\bibfnamefont {A.}~\bibnamefont
  {Esposito}}, \bibinfo {author} {\bibfnamefont {A.}~\bibnamefont {Pilloni}}, \
  and\ \bibinfo {author} {\bibfnamefont {A.~D.}\ \bibnamefont {Polosa}},\
  }\href {\doibase 10.1016/j.physrep.2016.11.002} {\bibfield  {journal}
  {\bibinfo  {journal} {Phys.Rept.}\ }\textbf {\bibinfo {volume} {668}},\
  \bibinfo {pages} {1} (\bibinfo {year} {2017})},\ \Eprint
  {http://arxiv.org/abs/1611.07920}{\tt arXiv:1611.07920 [hep-ph]}\BibitemShut
  {NoStop}%
\bibitem [{\citenamefont {Lebed}\ \emph {et~al.}(2017)\citenamefont {Lebed},
  \citenamefont {Mitchell},\ and\ \citenamefont {Swanson}}]{Lebed:2016hpi}%
  \BibitemOpen
  \bibfield  {author} {\bibinfo {author} {\bibfnamefont {R.~F.}\ \bibnamefont
  {Lebed}}, \bibinfo {author} {\bibfnamefont {R.~E.}\ \bibnamefont {Mitchell}},
  \ and\ \bibinfo {author} {\bibfnamefont {E.~S.}\ \bibnamefont {Swanson}},\
  }\href {\doibase 10.1016/j.ppnp.2016.11.003} {\bibfield  {journal} {\bibinfo
  {journal} {Prog.Part.Nucl.Phys.}\ }\textbf {\bibinfo {volume} {93}},\
  \bibinfo {pages} {143} (\bibinfo {year} {2017})},\ \Eprint
  {http://arxiv.org/abs/1610.04528}{\tt arXiv:1610.04528 [hep-ph]}\BibitemShut
  {NoStop}%
\bibitem [{\citenamefont {Guo}\ \emph {et~al.}(2018)\citenamefont {Guo},
  \citenamefont {Hanhart}, \citenamefont {Mei{\ss}ner}, \citenamefont {Wang},
  \citenamefont {Zhao},\ and\ \citenamefont {Zou}}]{Guo:2017jvc}%
  \BibitemOpen
  \bibfield  {author} {\bibinfo {author} {\bibfnamefont {F.-K.}\ \bibnamefont
  {Guo}}, \bibinfo {author} {\bibfnamefont {C.}~\bibnamefont {Hanhart}},
  \bibinfo {author} {\bibfnamefont {U.-G.}\ \bibnamefont {Mei{\ss}ner}},
  \bibinfo {author} {\bibfnamefont {Q.}~\bibnamefont {Wang}}, \bibinfo {author}
  {\bibfnamefont {Q.}~\bibnamefont {Zhao}}, \ and\ \bibinfo {author}
  {\bibfnamefont {B.-S.}\ \bibnamefont {Zou}},\ }\href {\doibase
  10.1103/RevModPhys.90.015004} {\bibfield  {journal} {\bibinfo  {journal}
  {Rev.Mod.Phys.}\ }\textbf {\bibinfo {volume} {90}},\ \bibinfo {pages}
  {015004} (\bibinfo {year} {2018})},\ \Eprint
  {http://arxiv.org/abs/1705.00141}{\tt arXiv:1705.00141 [hep-ph]}\BibitemShut
  {NoStop}%
\bibitem [{\citenamefont {Olsen}\ \emph {et~al.}(2018)\citenamefont {Olsen},
  \citenamefont {Skwarnicki},\ and\ \citenamefont {Zieminska}}]{Olsen:2017bmm}%
  \BibitemOpen
  \bibfield  {author} {\bibinfo {author} {\bibfnamefont {S.~L.}\ \bibnamefont
  {Olsen}}, \bibinfo {author} {\bibfnamefont {T.}~\bibnamefont {Skwarnicki}}, \
  and\ \bibinfo {author} {\bibfnamefont {D.}~\bibnamefont {Zieminska}},\ }\href
  {\doibase 10.1103/RevModPhys.90.015003} {\bibfield  {journal} {\bibinfo
  {journal} {Rev.Mod.Phys.}\ }\textbf {\bibinfo {volume} {90}},\ \bibinfo
  {pages} {015003} (\bibinfo {year} {2018})},\ \Eprint
  {http://arxiv.org/abs/1708.04012}{\tt arXiv:1708.04012 [hep-ph]}\BibitemShut
  {NoStop}%
\bibitem [{\citenamefont {Karliner}\ \emph {et~al.}(2018)\citenamefont
  {Karliner}, \citenamefont {Rosner},\ and\ \citenamefont
  {Skwarnicki}}]{Karliner:2017qhf}%
  \BibitemOpen
  \bibfield  {author} {\bibinfo {author} {\bibfnamefont {M.}~\bibnamefont
  {Karliner}}, \bibinfo {author} {\bibfnamefont {J.~L.}\ \bibnamefont
  {Rosner}}, \ and\ \bibinfo {author} {\bibfnamefont {T.}~\bibnamefont
  {Skwarnicki}},\ }\href {\doibase 10.1146/annurev-nucl-101917-020902}
  {\bibfield  {journal} {\bibinfo  {journal} {Ann.Rev.Nucl.Part.Sci}\ }\textbf
  {\bibinfo {volume} {68}} (\bibinfo {year} {2018}),\
  10.1146/annurev-nucl-101917-020902},\ \Eprint
  {http://arxiv.org/abs/1711.10626}{\tt arXiv:1711.10626 [hep-ph]}\BibitemShut
  {NoStop}%
\bibitem [{\citenamefont {Horn}\ and\ \citenamefont
  {Mandula}(1978)}]{Horn:1977rq}%
  \BibitemOpen
  \bibfield  {author} {\bibinfo {author} {\bibfnamefont {D.}~\bibnamefont
  {Horn}}\ and\ \bibinfo {author} {\bibfnamefont {J.}~\bibnamefont {Mandula}},\
  }\href {\doibase 10.1103/PhysRevD.17.898} {\bibfield  {journal} {\bibinfo
  {journal} {Phys.Rev.}\ }\textbf {\bibinfo {volume} {D17}},\ \bibinfo {pages}
  {898} (\bibinfo {year} {1978})}\BibitemShut {NoStop}%
\bibitem [{\citenamefont {Isgur}\ and\ \citenamefont
  {Paton}(1985)}]{Isgur:1984bm}%
  \BibitemOpen
  \bibfield  {author} {\bibinfo {author} {\bibfnamefont {N.}~\bibnamefont
  {Isgur}}\ and\ \bibinfo {author} {\bibfnamefont {J.~E.}\ \bibnamefont
  {Paton}},\ }\href {\doibase 10.1103/PhysRevD.31.2910} {\bibfield  {journal}
  {\bibinfo  {journal} {Phys.Rev.}\ }\textbf {\bibinfo {volume} {D31}},\
  \bibinfo {pages} {2910} (\bibinfo {year} {1985})}\BibitemShut {NoStop}%
\bibitem [{\citenamefont {Chanowitz}\ and\ \citenamefont
  {Sharpe}(1983)}]{Chanowitz:1982qj}%
  \BibitemOpen
  \bibfield  {author} {\bibinfo {author} {\bibfnamefont {M.~S.}\ \bibnamefont
  {Chanowitz}}\ and\ \bibinfo {author} {\bibfnamefont {S.~R.}\ \bibnamefont
  {Sharpe}},\ }\href {\doibase 10.1016/0550-3213(83)90561-8} {\bibfield
  {journal} {\bibinfo  {journal} {Nucl.Phys.}\ }\textbf {\bibinfo {volume}
  {B222}},\ \bibinfo {pages} {211} (\bibinfo {year} {1983})},\ \bibinfo {note}
  {[Erratum: Nucl. Phys.B228,588(1983)]}\BibitemShut {NoStop}%
\bibitem [{\citenamefont {Barnes}\ \emph {et~al.}(1983)\citenamefont {Barnes},
  \citenamefont {Close}, \citenamefont {de~Viron},\ and\ \citenamefont
  {Weyers}}]{Barnes:1982tx}%
  \BibitemOpen
  \bibfield  {author} {\bibinfo {author} {\bibfnamefont {T.}~\bibnamefont
  {Barnes}}, \bibinfo {author} {\bibfnamefont {F.}~\bibnamefont {Close}},
  \bibinfo {author} {\bibfnamefont {F.}~\bibnamefont {de~Viron}}, \ and\
  \bibinfo {author} {\bibfnamefont {J.}~\bibnamefont {Weyers}},\ }\href
  {\doibase 10.1016/0550-3213(83)90004-4} {\bibfield  {journal} {\bibinfo
  {journal} {Nucl.Phys.}\ }\textbf {\bibinfo {volume} {B224}},\ \bibinfo
  {pages} {241} (\bibinfo {year} {1983})}\BibitemShut {NoStop}%
\bibitem [{\citenamefont {Close}\ and\ \citenamefont
  {Page}(1995)}]{Close:1994hc}%
  \BibitemOpen
  \bibfield  {author} {\bibinfo {author} {\bibfnamefont {F.~E.}\ \bibnamefont
  {Close}}\ and\ \bibinfo {author} {\bibfnamefont {P.~R.}\ \bibnamefont
  {Page}},\ }\href {\doibase 10.1016/0550-3213(95)00085-7} {\bibfield
  {journal} {\bibinfo  {journal} {Nucl.Phys.}\ }\textbf {\bibinfo {volume}
  {B443}},\ \bibinfo {pages} {233} (\bibinfo {year} {1995})},\ \Eprint
  {http://arxiv.org/abs/hep-ph/9411301}{\tt arXiv:hep-ph/9411301
  [hep-ph]}\BibitemShut {NoStop}%
\bibitem [{\citenamefont {Lacock}\ \emph {et~al.}(1997)\citenamefont {Lacock},
  \citenamefont {Michael}, \citenamefont {Boyle},\ and\ \citenamefont
  {Rowland}}]{Lacock:1996ny}%
  \BibitemOpen
  \bibfield  {author} {\bibinfo {author} {\bibfnamefont {P.}~\bibnamefont
  {Lacock}}, \bibinfo {author} {\bibfnamefont {C.}~\bibnamefont {Michael}},
  \bibinfo {author} {\bibfnamefont {P.}~\bibnamefont {Boyle}}, \ and\ \bibinfo
  {author} {\bibfnamefont {P.}~\bibnamefont {Rowland}} (\bibinfo
  {collaboration} {UKQCD} Collaboration),\ }\href {\doibase
  10.1016/S0370-2693(97)00384-5} {\bibfield  {journal} {\bibinfo  {journal}
  {Phys.Lett.}\ }\textbf {\bibinfo {volume} {B401}},\ \bibinfo {pages} {308}
  (\bibinfo {year} {1997})},\ \Eprint
  {http://arxiv.org/abs/hep-lat/9611011}{\tt arXiv:hep-lat/9611011
  [hep-lat]}\BibitemShut {NoStop}%
\bibitem [{\citenamefont {Bernard}\ \emph {et~al.}(1997)\citenamefont {Bernard}
  \emph {et~al.}}]{Bernard:1997ib}%
  \BibitemOpen
  \bibfield  {author} {\bibinfo {author} {\bibfnamefont {C.~W.}\ \bibnamefont
  {Bernard}} \emph {et~al.} (\bibinfo {collaboration} {MILC} Collaboration),\
  }\href {\doibase 10.1103/PhysRevD.56.7039} {\bibfield  {journal} {\bibinfo
  {journal} {Phys.Rev.}\ }\textbf {\bibinfo {volume} {D56}},\ \bibinfo {pages}
  {7039} (\bibinfo {year} {1997})},\ \Eprint
  {http://arxiv.org/abs/hep-lat/9707008}{\tt arXiv:hep-lat/9707008
  [hep-lat]}\BibitemShut {NoStop}%
\bibitem [{\citenamefont {Dudek}\ \emph {et~al.}(2013)\citenamefont {Dudek},
  \citenamefont {Edwards}, \citenamefont {Guo},\ and\ \citenamefont
  {Thomas}}]{Dudek:2013yja}%
  \BibitemOpen
  \bibfield  {author} {\bibinfo {author} {\bibfnamefont {J.~J.}\ \bibnamefont
  {Dudek}}, \bibinfo {author} {\bibfnamefont {R.~G.}\ \bibnamefont {Edwards}},
  \bibinfo {author} {\bibfnamefont {P.}~\bibnamefont {Guo}}, \ and\ \bibinfo
  {author} {\bibfnamefont {C.~E.}\ \bibnamefont {Thomas}} (\bibinfo
  {collaboration} {Hadron Spectrum} Collaboration),\ }\href {\doibase
  10.1103/PhysRevD.88.094505} {\bibfield  {journal} {\bibinfo  {journal}
  {Phys.Rev.}\ }\textbf {\bibinfo {volume} {D88}},\ \bibinfo {pages} {094505}
  (\bibinfo {year} {2013})},\ \Eprint {http://arxiv.org/abs/1309.2608}{\tt
  arXiv:1309.2608 [hep-lat]}\BibitemShut {NoStop}%
\bibitem [{\citenamefont {Szczepaniak}\ and\ \citenamefont
  {Swanson}(2002)}]{Szczepaniak:2001rg}%
  \BibitemOpen
  \bibfield  {author} {\bibinfo {author} {\bibfnamefont {A.~P.}\ \bibnamefont
  {Szczepaniak}}\ and\ \bibinfo {author} {\bibfnamefont {E.~S.}\ \bibnamefont
  {Swanson}},\ }\href {\doibase 10.1103/PhysRevD.65.025012} {\bibfield
  {journal} {\bibinfo  {journal} {Phys.Rev.}\ }\textbf {\bibinfo {volume}
  {D65}},\ \bibinfo {pages} {025012} (\bibinfo {year} {2002})},\ \Eprint
  {http://arxiv.org/abs/hep-ph/0107078}{\tt arXiv:hep-ph/0107078
  [hep-ph]}\BibitemShut {NoStop}%
\bibitem [{\citenamefont {Szczepaniak}\ and\ \citenamefont
  {Krupinski}(2006)}]{Szczepaniak:2006nx}%
  \BibitemOpen
  \bibfield  {author} {\bibinfo {author} {\bibfnamefont {A.~P.}\ \bibnamefont
  {Szczepaniak}}\ and\ \bibinfo {author} {\bibfnamefont {P.}~\bibnamefont
  {Krupinski}},\ }\href {\doibase 10.1103/PhysRevD.73.116002} {\bibfield
  {journal} {\bibinfo  {journal} {Phys.Rev.}\ }\textbf {\bibinfo {volume}
  {D73}},\ \bibinfo {pages} {116002} (\bibinfo {year} {2006})},\ \Eprint
  {http://arxiv.org/abs/hep-ph/0604098}{\tt arXiv:hep-ph/0604098
  [hep-ph]}\BibitemShut {NoStop}%
\bibitem [{\citenamefont {Guo}\ \emph {et~al.}(2008)\citenamefont {Guo},
  \citenamefont {Szczepaniak}, \citenamefont {Galata}, \citenamefont
  {Vassallo},\ and\ \citenamefont {Santopinto}}]{Guo:2008yz}%
  \BibitemOpen
  \bibfield  {author} {\bibinfo {author} {\bibfnamefont {P.}~\bibnamefont
  {Guo}}, \bibinfo {author} {\bibfnamefont {A.~P.}\ \bibnamefont
  {Szczepaniak}}, \bibinfo {author} {\bibfnamefont {G.}~\bibnamefont {Galata}},
  \bibinfo {author} {\bibfnamefont {A.}~\bibnamefont {Vassallo}}, \ and\
  \bibinfo {author} {\bibfnamefont {E.}~\bibnamefont {Santopinto}},\ }\href
  {\doibase 10.1103/PhysRevD.78.056003} {\bibfield  {journal} {\bibinfo
  {journal} {Phys.Rev.}\ }\textbf {\bibinfo {volume} {D78}},\ \bibinfo {pages}
  {056003} (\bibinfo {year} {2008})},\ \Eprint
  {http://arxiv.org/abs/0807.2721}{\tt arXiv:0807.2721 [hep-ph]}\BibitemShut
  {NoStop}%
\bibitem [{\citenamefont {Bass}\ and\ \citenamefont
  {Moskal}(2018)}]{Bass:2018xmz}%
  \BibitemOpen
  \bibfield  {author} {\bibinfo {author} {\bibfnamefont {S.~D.}\ \bibnamefont
  {Bass}}\ and\ \bibinfo {author} {\bibfnamefont {P.}~\bibnamefont {Moskal}},\
  }\href@noop {} {} (\bibinfo {year} {2018}),\ \Eprint
  {http://arxiv.org/abs/1810.12290}{\tt arXiv:1810.12290 [hep-ph]}\BibitemShut
  {NoStop}%
\bibitem [{\citenamefont {Baum}\ \emph {et~al.}(1996)\citenamefont {Baum} \emph
  {et~al.}}]{Baum:1996yv}%
  \BibitemOpen
  \bibfield  {author} {\bibinfo {author} {\bibfnamefont {G.}~\bibnamefont
  {Baum}} \emph {et~al.} (\bibinfo {collaboration} {COMPASS} Collaboration),\
  }\emph {\bibinfo {title}{COMPASS: A Proposal for a Common Muon and Proton Apparatus for Structure and Spectroscopy}} \href@noop {} {} (\bibinfo {year} {1996}), \bibinfo {note} {\href{http://cds.cern.ch/record/298433}{CERN-SPSLC-96-14}}\BibitemShut {NoStop}%
\bibitem [{\citenamefont {Abbon}\ \emph {et~al.}(2015)\citenamefont {Abbon}
  \emph {et~al.}}]{Abbon:2014aex}%
  \BibitemOpen
  \bibfield  {author} {\bibinfo {author} {\bibfnamefont {P.}~\bibnamefont
  {Abbon}} \emph {et~al.} (\bibinfo {collaboration} {COMPASS} Collaboration),\
  }\href {\doibase 10.1016/j.nima.2015.01.035} {\bibfield  {journal} {\bibinfo
  {journal} {Nucl.Instrum.Meth.}\ }\textbf {\bibinfo {volume} {A779}},\
  \bibinfo {pages} {68} (\bibinfo {year} {2015})},\ \Eprint
  {http://arxiv.org/abs/1410.1797}{\tt arXiv:1410.1797
  [physics.ins-det]}\BibitemShut {NoStop}%
\bibitem [{\citenamefont {Rizzo}(2016)}]{Rizzo:2016qvl}%
  \BibitemOpen
  \bibfield  {author} {\bibinfo {author} {\bibfnamefont {A.}~\bibnamefont
  {Rizzo}} (\bibinfo {collaboration} {CLAS} Collaboration),\ }\href {\doibase
  10.1088/1742-6596/689/1/012022} {\bibfield  {journal} {\bibinfo  {journal}
  {J.Phys.Conf.Ser.}\ }\textbf {\bibinfo {volume} {689}},\ \bibinfo {pages}
  {012022} (\bibinfo {year} {2016})}\BibitemShut {NoStop}%
\bibitem [{\citenamefont {Dobbs}(2018)}]{Dobbs:2017vjw}%
  \BibitemOpen
  \bibfield  {author} {\bibinfo {author} {\bibfnamefont {S.}~\bibnamefont
  {Dobbs}} (\bibinfo {collaboration} {GlueX} Collaboration),\ }\href {\doibase
  10.22323/1.310.0047} {\bibfield  {journal} {\bibinfo  {journal} {PoS}\
  }\textbf {\bibinfo {volume} {Hadron2017}},\ \bibinfo {pages} {047} (\bibinfo
  {year} {2018})},\ \Eprint {http://arxiv.org/abs/1712.07214}{\tt
  arXiv:1712.07214 [nucl-ex]}\BibitemShut {NoStop}%
\bibitem [{\citenamefont {Thompson}\ \emph {et~al.}(1997)\citenamefont
  {Thompson} \emph {et~al.}}]{Thompson:1997bs}%
  \BibitemOpen
  \bibfield  {author} {\bibinfo {author} {\bibfnamefont {D.~R.}\ \bibnamefont
  {Thompson}} \emph {et~al.} (\bibinfo {collaboration} {E852} Collaboration),\
  }\href {\doibase 10.1103/PhysRevLett.79.1630} {\bibfield  {journal} {\bibinfo
   {journal} {Phys.Rev.Lett.}\ }\textbf {\bibinfo {volume} {79}},\ \bibinfo
  {pages} {1630} (\bibinfo {year} {1997})},\ \Eprint
  {http://arxiv.org/abs/hep-ex/9705011}{\tt arXiv:hep-ex/9705011
  [hep-ex]}\BibitemShut {NoStop}%
\bibitem [{\citenamefont {Chung}\ \emph {et~al.}(1999)\citenamefont {Chung}
  \emph {et~al.}}]{Chung:1999we}%
  \BibitemOpen
  \bibfield  {author} {\bibinfo {author} {\bibfnamefont {S.~U.}\ \bibnamefont
  {Chung}} \emph {et~al.} (\bibinfo {collaboration} {E852} Collaboration),\
  }\href {\doibase 10.1103/PhysRevD.60.092001} {\bibfield  {journal} {\bibinfo
  {journal} {Phys.Rev.}\ }\textbf {\bibinfo {volume} {D60}},\ \bibinfo {pages}
  {092001} (\bibinfo {year} {1999})},\ \Eprint
  {http://arxiv.org/abs/hep-ex/9902003}{\tt arXiv:hep-ex/9902003
  [hep-ex]}\BibitemShut {NoStop}%
\bibitem [{\citenamefont {Adams}\ \emph {et~al.}(2007)\citenamefont {Adams}
  \emph {et~al.}}]{Adams:2006sa}%
  \BibitemOpen
  \bibfield  {author} {\bibinfo {author} {\bibfnamefont {G.~S.}\ \bibnamefont
  {Adams}} \emph {et~al.} (\bibinfo {collaboration} {E852} Collaboration),\
  }\href {\doibase 10.1016/j.physletb.2007.07.068} {\bibfield  {journal}
  {\bibinfo  {journal} {Phys.Lett.}\ }\textbf {\bibinfo {volume} {B657}},\
  \bibinfo {pages} {27} (\bibinfo {year} {2007})},\ \Eprint
  {http://arxiv.org/abs/hep-ex/0612062}{\tt arXiv:hep-ex/0612062
  [hep-ex]}\BibitemShut {NoStop}%
\bibitem [{\citenamefont {Abele}\ \emph {et~al.}(1998)\citenamefont {Abele}
  \emph {et~al.}}]{Abele:1998gn}%
  \BibitemOpen
  \bibfield  {author} {\bibinfo {author} {\bibfnamefont {A.}~\bibnamefont
  {Abele}} \emph {et~al.} (\bibinfo {collaboration} {C
rystal Barrel}
  Collaboration),\ }\href {\doibase 10.1016/S0370-2693(98)00123-3} {\bibfield
  {journal} {\bibinfo  {journal} {Phys.Lett.}\ }\textbf {\bibinfo {volume}
  {B423}},\ \bibinfo {pages} {175} (\bibinfo {year} {1998})}\BibitemShut
  {NoStop}%
\bibitem [{\citenamefont {Abele}\ \emph {et~al.}(1999)\citenamefont {Abele}
  \emph {et~al.}}]{Abele:1999tf}%
  \BibitemOpen
  \bibfield  {author} {\bibinfo {author} {\bibfnamefont {A.}~\bibnamefont
  {Abele}} \emph {et~al.} (\bibinfo {collaboration} {Crystal Barrel}
  Collaboration),\ }\href {\doibase 10.1016/S0370-2693(98)01544-5} {\bibfield
  {journal} {\bibinfo  {journal} {Phys.Lett.}\ }\textbf {\bibinfo {volume}
  {B446}},\ \bibinfo {pages} {349} (\bibinfo {year} {1999})}\BibitemShut
  {NoStop}%
\bibitem [{\citenamefont {Ivanov}\ \emph {et~al.}(2001)\citenamefont {Ivanov}
  \emph {et~al.}}]{Ivanov:2001rv}%
  \BibitemOpen
  \bibfield  {author} {\bibinfo {author} {\bibfnamefont {E.~I.}\ \bibnamefont
  {Ivanov}} \emph {et~al.} (\bibinfo {collaboration} {E852} Collaboration),\
  }\href {\doibase 10.1103/PhysRevLett.86.3977} {\bibfield  {journal} {\bibinfo
   {journal} {Phys.Rev.Lett.}\ }\textbf {\bibinfo {volume} {86}},\ \bibinfo
  {pages} {3977} (\bibinfo {year} {2001})},\ \Eprint
  {http://arxiv.org/abs/hep-ex/0101058}{\tt arXiv:hep-ex/0101058
  [hep-ex]}\BibitemShut {NoStop}%
\bibitem [{\citenamefont {Khokhlov}(2000)}]{Khokhlov:2000tk}%
  \BibitemOpen
  \bibfield  {author} {\bibinfo {author} {\bibfnamefont {{\relax Yu}.~A.}\
  \bibnamefont {Khokhlov}} (\bibinfo {collaboration} {VES} Collaboration),\
  }\href {\doibase 10.1016/S0375-9474(99)00663-6} {\bibfield  {journal}
  {\bibinfo  {journal} {Nucl.Phys.}\ }\textbf {\bibinfo {volume} {A663}},\
  \bibinfo {pages} {596} (\bibinfo {year} {2000})},\ \bibinfo {note}
  {{Particles and nuclei. Proceedings, 15th International Conference, PANIC
  '99, Uppsala, Sweden, June 10-16, 1999}}\BibitemShut {NoStop}%
\bibitem [{\citenamefont {Alekseev}\ \emph {et~al.}(2010)\citenamefont
  {Alekseev} \emph {et~al.}}]{Alekseev:2009aa}%
  \BibitemOpen
  \bibfield  {author} {\bibinfo {author} {\bibfnamefont {M.}~\bibnamefont
  {Alekseev}} \emph {et~al.} (\bibinfo {collaboration} {COMPASS}
  Collaboration),\ }\href {\doibase 10.1103/PhysRevLett.104.241803} {\bibfield
  {journal} {\bibinfo  {journal} {Phys.Rev.Lett.}\ }\textbf {\bibinfo {volume}
  {104}},\ \bibinfo {pages} {241803} (\bibinfo {year} {2010})},\ \Eprint
  {http://arxiv.org/abs/0910.5842}{\tt arXiv:0910.5842 [hep-ex]}\BibitemShut
  {NoStop}%
\bibitem [{\citenamefont {Akhunzyanov}\ \emph {et~al.}(2018)\citenamefont
  {Akhunzyanov} \emph {et~al.}}]{Akhunzyanov:2018pnr}%
  \BibitemOpen
  \bibfield  {author} {\bibinfo {author} {\bibfnamefont {R.}~\bibnamefont
  {Akhunzyanov}} \emph {et~al.} (\bibinfo {collaboration} {COMPASS}
  Collaboration),\ }\href@noop {} {} (\bibinfo {year} {2018}),\ \Eprint
  {http://arxiv.org/abs/1802.05913}{\tt arXiv:1802.05913 [hep-ex]}\BibitemShut
  {NoStop}%
\bibitem [{\citenamefont {Adolph}\ \emph {et~al.}(2015)\citenamefont {Adolph}
  \emph {et~al.}}]{Adolph:2014rpp}%
  \BibitemOpen
  \bibfield  {author} {\bibinfo {author} {\bibfnamefont {C.}~\bibnamefont
  {Adolph}} \emph {et~al.} (\bibinfo {collaboration} {COMPASS} Collaboration),\
  }\href {\doibase 10.1016/j.physletb.2014.11.058} {\bibfield  {journal}
  {\bibinfo  {journal} {Phys.Lett.}\ }\textbf {\bibinfo {volume} {B740}},\
  \bibinfo {pages} {303} (\bibinfo {year} {2015})},\ \Eprint
  {http://arxiv.org/abs/1408.4286}{\tt arXiv:1408.4286 [hep-ex]}\BibitemShut
  {NoStop}%
\bibitem [{\citenamefont {Bass}\ and\ \citenamefont
  {Marco}(2002)}]{Bass:2001zs}%
  \BibitemOpen
  \bibfield  {author} {\bibinfo {author} {\bibfnamefont {S.~D.}\ \bibnamefont
  {Bass}}\ and\ \bibinfo {author} {\bibfnamefont {E.}~\bibnamefont {Marco}},\
  }\href {\doibase 10.1103/PhysRevD.65.057503} {\bibfield  {journal} {\bibinfo
  {journal} {Phys.Rev.}\ }\textbf {\bibinfo {volume} {D65}},\ \bibinfo {pages}
  {057503} (\bibinfo {year} {2002})},\ \Eprint
  {http://arxiv.org/abs/hep-ph/0108189}{\tt arXiv:hep-ph/0108189
  [hep-ph]}\BibitemShut {NoStop}%
\bibitem [{\citenamefont {Szczepaniak}\ \emph {et~al.}(2003)\citenamefont
  {Szczepaniak}, \citenamefont {Swat}, \citenamefont {Dzierba},\ and\
  \citenamefont {Teige}}]{Szczepaniak:2003vg}%
  \BibitemOpen
  \bibfield  {author} {\bibinfo {author} {\bibfnamefont {A.~P.}\ \bibnamefont
  {Szczepaniak}}, \bibinfo {author} {\bibfnamefont {M.}~\bibnamefont {Swat}},
  \bibinfo {author} {\bibfnamefont {A.~R.}\ \bibnamefont {Dzierba}}, \ and\
  \bibinfo {author} {\bibfnamefont {S.}~\bibnamefont {Teige}},\ }\href
  {\doibase 10.1103/PhysRevLett.91.092002} {\bibfield  {journal} {\bibinfo
  {journal} {Phys.Rev.Lett.}\ }\textbf {\bibinfo {volume} {91}},\ \bibinfo
  {pages} {092002} (\bibinfo {year} {2003})},\ \Eprint
  {http://arxiv.org/abs/hep-ph/0304095}{\tt arXiv:hep-ph/0304095
  [hep-ph]}\BibitemShut {NoStop}%
\bibitem [{\citenamefont {Close}\ and\ \citenamefont
  {Lipkin}(1987)}]{Close:1987aw}%
  \BibitemOpen
  \bibfield  {author} {\bibinfo {author} {\bibfnamefont {F.~E.}\ \bibnamefont
  {Close}}\ and\ \bibinfo {author} {\bibfnamefont {H.~J.}\ \bibnamefont
  {Lipkin}},\ }\href {\doibase 10.1016/0370-2693(87)90613-7} {\bibfield
  {journal} {\bibinfo  {journal} {Phys.Lett.}\ }\textbf {\bibinfo {volume}
  {B196}},\ \bibinfo {pages} {245} (\bibinfo {year} {1987})}\BibitemShut
  {NoStop}%
\bibitem [{\citenamefont {Chung}\ \emph {et~al.}(2002)\citenamefont {Chung},
  \citenamefont {Klempt},\ and\ \citenamefont {Korner}}]{Chung:2002fz}%
  \BibitemOpen
  \bibfield  {author} {\bibinfo {author} {\bibfnamefont {S.~U.}\ \bibnamefont
  {Chung}}, \bibinfo {author} {\bibfnamefont {E.}~\bibnamefont {Klempt}}, \
  and\ \bibinfo {author} {\bibfnamefont {J.~G.}\ \bibnamefont {Korner}},\
  }\href {\doibase 10.1140/epja/i2002-10058-0} {\bibfield  {journal} {\bibinfo
  {journal} {Eur.Phys.J.}\ }\textbf {\bibinfo {volume} {A15}},\ \bibinfo
  {pages} {539} (\bibinfo {year} {2002})},\ \Eprint
  {http://arxiv.org/abs/hep-ph/0211100}{\tt arXiv:hep-ph/0211100
  [hep-ph]}\BibitemShut {NoStop}%
\bibitem [{\citenamefont {Jaffe}\ and\ \citenamefont
  {Wilczek}(2003)}]{Jaffe:2003sg}%
  \BibitemOpen
  \bibfield  {author} {\bibinfo {author} {\bibfnamefont {R.~L.}\ \bibnamefont
  {Jaffe}}\ and\ \bibinfo {author} {\bibfnamefont {F.}~\bibnamefont
  {Wilczek}},\ }\href {\doibase 10.1103/PhysRevLett.91.232003} {\bibfield
  {journal} {\bibinfo  {journal} {Phys.Rev.Lett.}\ }\textbf {\bibinfo {volume}
  {91}},\ \bibinfo {pages} {232003} (\bibinfo {year} {2003})},\ \Eprint
  {http://arxiv.org/abs/hep-ph/0307341}{\tt arXiv:hep-ph/0307341
  [hep-ph]}\BibitemShut {NoStop}%
\bibitem [{\citenamefont {Jaffe}(2005)}]{Jaffe:2004ph}%
  \BibitemOpen
  \bibfield  {author} {\bibinfo {author} {\bibfnamefont {R.}~\bibnamefont
  {Jaffe}},\ }\href {\doibase 10.1016/j.physrep.2004.11.005} {\bibfield
  {journal} {\bibinfo  {journal} {Phys.Rept.}\ }\textbf {\bibinfo {volume}
  {409}},\ \bibinfo {pages} {1} (\bibinfo {year} {2005})},\ \Eprint
  {http://arxiv.org/abs/hep-ph/0409065}{\tt arXiv:hep-ph/0409065
  [hep-ph]}\BibitemShut {NoStop}%
\bibitem [{\citenamefont {Jackura}\ \emph {et~al.}(2017)\citenamefont {Jackura}
  \emph {et~al.}}]{Jackura:2017amb}%
  \BibitemOpen
  \bibfield  {author} {\bibinfo {author} {\bibfnamefont {A.}~\bibnamefont
  {Jackura}} \emph {et~al.} (\bibinfo {collaboration} {COMPASS and JPAC}
  Collaborations),\ }\href {\doibase
  https://doi.org/10.1016/j.physletb.2018.01.017} {\bibfield  {journal}
  {\bibinfo  {journal} {Phys.Lett.}\ }\textbf {\bibinfo {volume} {B779}},\
  \bibinfo {pages} {464–472} (\bibinfo {year} {2017})},\ \Eprint
  {http://arxiv.org/abs/1707.02848}{\tt arXiv:1707.02848 [hep-ph]}\BibitemShut
  {NoStop}%
\bibitem [{\citenamefont {Schl\"uter}(2012)}]{Schluter:2012mep}%
  \BibitemOpen
  \bibfield  {author} {\bibinfo {author} {\bibfnamefont {T.}~\bibnamefont
  {Schl\"uter}},\ }\emph {\bibinfo {title} {The $\pi^-\eta$ and
  $\pi^-\eta^\prime$ Systems in Exclusive 190~GeV $\pi^-p$ Reactions at
  COMPASS}},\ \href@noop {} {Ph.D. thesis},\ \bibinfo  {school} {Munich U.}
  (\bibinfo {year} {2012})\BibitemShut {NoStop}%
\bibitem [{\citenamefont {Kuhn}\ \emph {et~al.}(2004)\citenamefont {Kuhn} \emph
  {et~al.}}]{Kuhn:2004en}%
  \BibitemOpen
  \bibfield  {author} {\bibinfo {author} {\bibfnamefont {J.}~\bibnamefont
  {Kuhn}} \emph {et~al.} (\bibinfo {collaboration} {E852} Collaboration),\
  }\href {\doibase 10.1016/j.physletb.2004.05.032} {\bibfield  {journal}
  {\bibinfo  {journal} {Phys.Lett.}\ }\textbf {\bibinfo {volume} {B595}},\
  \bibinfo {pages} {109} (\bibinfo {year} {2004})},\ \Eprint
  {http://arxiv.org/abs/hep-ex/0401004}{\tt arXiv:hep-ex/0401004
  [hep-ex]}\BibitemShut {NoStop}%
\bibitem [{\citenamefont {Lu}\ \emph {et~al.}(2005)\citenamefont {Lu} \emph
  {et~al.}}]{Lu:2004yn}%
  \BibitemOpen
  \bibfield  {author} {\bibinfo {author} {\bibfnamefont {M.}~\bibnamefont {Lu}}
  \emph {et~al.} (\bibinfo {collaboration} {E852} Collaboration),\ }\href
  {\doibase 10.1103/PhysRevLett.94.032002} {\bibfield  {journal} {\bibinfo
  {journal} {Phys.Rev.Lett.}\ }\textbf {\bibinfo {volume} {94}},\ \bibinfo
  {pages} {032002} (\bibinfo {year} {2005})},\ \Eprint
  {http://arxiv.org/abs/hep-ex/0405044}{\tt arXiv:hep-ex/0405044
  [hep-ex]}\BibitemShut {NoStop}%
\bibitem [{\citenamefont {Deck}(1964)}]{Deck:1964hm}%
  \BibitemOpen
  \bibfield  {author} {\bibinfo {author} {\bibfnamefont {R.~T.}\ \bibnamefont
  {Deck}},\ }\href {\doibase 10.1103/PhysRevLett.13.169} {\bibfield  {journal}
  {\bibinfo  {journal} {Phys.Rev.Lett.}\ }\textbf {\bibinfo {volume} {13}},\
  \bibinfo {pages} {169} (\bibinfo {year} {1964})}\BibitemShut {NoStop}%
\bibitem [{\citenamefont {Ascoli}\ \emph {et~al.}(1974)\citenamefont {Ascoli},
  \citenamefont {Cutler}, \citenamefont {Jones}, \citenamefont {Kruse},
  \citenamefont {Roberts}, \citenamefont {Weinstein},\ and\ \citenamefont
  {Wyld}}]{Ascoli:1974hi}%
  \BibitemOpen
  \bibfield  {author} {\bibinfo {author} {\bibfnamefont {G.}~\bibnamefont
  {Ascoli}}, \bibinfo {author} {\bibfnamefont {R.}~\bibnamefont {Cutler}},
  \bibinfo {author} {\bibfnamefont {L.~M.}\ \bibnamefont {Jones}}, \bibinfo
  {author} {\bibfnamefont {U.}~\bibnamefont {Kruse}}, \bibinfo {author}
  {\bibfnamefont {T.}~\bibnamefont {Roberts}}, \bibinfo {author} {\bibfnamefont
  {B.}~\bibnamefont {Weinstein}}, \ and\ \bibinfo {author} {\bibfnamefont
  {H.~W.}\ \bibnamefont {Wyld}},\ }\href {\doibase 10.1103/PhysRevD.9.1963}
  {\bibfield  {journal} {\bibinfo  {journal} {Phys.Rev.}\ }\textbf {\bibinfo
  {volume} {D9}},\ \bibinfo {pages} {1963} (\bibinfo {year}
  {1974})}\BibitemShut {NoStop}%
\bibitem [{\citenamefont {Ryabchikov}()}]{dimadeck}%
  \BibitemOpen
  \bibfield  {author} {\bibinfo {author} {\bibfnamefont {D.}~\bibnamefont
  {Ryabchikov}},\ }\href@noop {} {}\bibinfo {note} {Talk at PWA9/ATHOS4,
  \href{https://indico.cern.ch/event/591374/contributions/2498368/attachments/1427728/2191292/03_ryabchikov_athos2017.pdf}{https://indico.cern.ch/event/591374/contributions/ 2498368/attachments/1427728/2191292/ 03\_ryabchikov\_athos2017.pdf}}\BibitemShut
  {NoStop}%
\bibitem [{\citenamefont {Chew}\ and\ \citenamefont
  {Frautschi}(1961)}]{Chew:1961ev}%
  \BibitemOpen
  \bibfield  {author} {\bibinfo {author} {\bibfnamefont {G.~F.}\ \bibnamefont
  {Chew}}\ and\ \bibinfo {author} {\bibfnamefont {S.~C.}\ \bibnamefont
  {Frautschi}},\ }\href {\doibase 10.1103/PhysRevLett.7.394} {\bibfield
  {journal} {\bibinfo  {journal} {Phys.Rev.Lett.}\ }\textbf {\bibinfo {volume}
  {7}},\ \bibinfo {pages} {394} (\bibinfo {year} {1961})}\BibitemShut {NoStop}%
\bibitem [{\citenamefont {Donnachie}\ \emph {et~al.}(2005)\citenamefont
  {Donnachie}, \citenamefont {Dosch}, \citenamefont {Nachtmann},\ and\
  \citenamefont {Landshoff}}]{Donnachie:2002en}%
  \BibitemOpen
  \bibfield  {author} {\bibinfo {author} {\bibfnamefont {S.}~\bibnamefont
  {Donnachie}}, \bibinfo {author} {\bibfnamefont {H.~G.}\ \bibnamefont
  {Dosch}}, \bibinfo {author} {\bibfnamefont {O.}~\bibnamefont {Nachtmann}}, \
  and\ \bibinfo {author} {\bibfnamefont {P.}~\bibnamefont {Landshoff}},\ }\href
  {https://books.google.it/books?id=WpGHPwAACAAJ} {\emph {\bibinfo {title}
  {{Pomeron physics and QCD}}}},\ Cambridge Monographs on Particle Physics,
  Nuclear Physics and Cosmology\ (\bibinfo  {publisher} {Cambridge University
  Press},\ \bibinfo {year} {2005})\BibitemShut {NoStop}%
\bibitem [{\citenamefont {Close}\ and\ \citenamefont
  {Schuler}(1999)}]{Close:1999is}%
  \BibitemOpen
  \bibfield  {author} {\bibinfo {author} {\bibfnamefont {F.~E.}\ \bibnamefont
  {Close}}\ and\ \bibinfo {author} {\bibfnamefont {G.~A.}\ \bibnamefont
  {Schuler}},\ }\href {\doibase 10.1016/S0370-2693(99)00450-5} {\bibfield
  {journal} {\bibinfo  {journal} {Phys.Lett.}\ }\textbf {\bibinfo {volume}
  {B458}},\ \bibinfo {pages} {127} (\bibinfo {year} {1999})},\ \Eprint
  {http://arxiv.org/abs/hep-ph/9902243}{\tt arXiv:hep-ph/9902243
  [hep-ph]}\BibitemShut {NoStop}%
\bibitem [{\citenamefont {Arens}\ \emph {et~al.}(1997)\citenamefont {Arens},
  \citenamefont {Nachtmann}, \citenamefont {Diehl},\ and\ \citenamefont
  {Landshoff}}]{Arens:1996xw}%
  \BibitemOpen
  \bibfield  {author} {\bibinfo {author} {\bibfnamefont {T.}~\bibnamefont
  {Arens}}, \bibinfo {author} {\bibfnamefont {O.}~\bibnamefont {Nachtmann}},
  \bibinfo {author} {\bibfnamefont {M.}~\bibnamefont {Diehl}}, \ and\ \bibinfo
  {author} {\bibfnamefont {P.~V.}\ \bibnamefont {Landshoff}},\ }\href {\doibase
  10.1007/s002880050430} {\bibfield  {journal} {\bibinfo  {journal} {Z.Phys.}\
  }\textbf {\bibinfo {volume} {C74}},\ \bibinfo {pages} {651} (\bibinfo {year}
  {1997})},\ \Eprint {http://arxiv.org/abs/hep-ph/9605376}{\tt
  arXiv:hep-ph/9605376 [hep-ph]}\BibitemShut {NoStop}%
\bibitem [{\citenamefont {Bjorken}(1960)}]{Bjorken:1960zz}%
  \BibitemOpen
  \bibfield  {author} {\bibinfo {author} {\bibfnamefont {J.~D.}\ \bibnamefont
  {Bjorken}},\ }\href {\doibase 10.1103/PhysRevLett.4.473} {\bibfield
  {journal} {\bibinfo  {journal} {Phys.Rev.Lett.}\ }\textbf {\bibinfo {volume}
  {4}},\ \bibinfo {pages} {473} (\bibinfo {year} {1960})}\BibitemShut {NoStop}%
\bibitem [{\citenamefont {Aitchison}(1972)}]{Aitchison:1972ay}%
  \BibitemOpen
  \bibfield  {author} {\bibinfo {author} {\bibfnamefont {I.~J.~R.}\
  \bibnamefont {Aitchison}},\ }\href {\doibase 10.1016/0375-9474(72)90305-3}
  {\bibfield  {journal} {\bibinfo  {journal} {Nucl.Phys.}\ }\textbf {\bibinfo
  {volume} {A189}},\ \bibinfo {pages} {417} (\bibinfo {year}
  {1972})}\BibitemShut {NoStop}%
\bibitem [{\citenamefont {James}\ and\ \citenamefont {Roos}(1975)}]{minuit}%
  \BibitemOpen
  \bibfield  {author} {\bibinfo {author} {\bibfnamefont {F.}~\bibnamefont
  {James}}\ and\ \bibinfo {author} {\bibfnamefont {M.}~\bibnamefont {Roos}},\
  }\href {\doibase 10.1016/0010-4655(75)90039-9} {\bibfield  {journal}
  {\bibinfo  {journal} {Comput.Phys.Commun.}\ }\textbf {\bibinfo {volume}
  {10}},\ \bibinfo {pages} {343} (\bibinfo {year} {1975})}\BibitemShut
  {NoStop}%
\bibitem [{\citenamefont {Press}\ \emph {et~al.}(2007)\citenamefont {Press},
  \citenamefont {Teukolsky}, \citenamefont {Vetterling},\ and\ \citenamefont
  {Flannery}}]{recipes}%
  \BibitemOpen
  \bibfield  {author} {\bibinfo {author} {\bibfnamefont {W.~H.}\ \bibnamefont
  {Press}}, \bibinfo {author} {\bibfnamefont {S.~A.}\ \bibnamefont
  {Teukolsky}}, \bibinfo {author} {\bibfnamefont {W.~T.}\ \bibnamefont
  {Vetterling}}, \ and\ \bibinfo {author} {\bibfnamefont {B.~P.}\ \bibnamefont
  {Flannery}},\ }\href@noop {} {\emph {\bibinfo {title} {Numerical Recipes 3rd
  Edition: The Art of Scientific Computing}}},\ \bibinfo {edition} {3rd}\ ed.\
  (\bibinfo  {publisher} {Cambridge University Press},\ \bibinfo {address} {New
  York, NY, USA},\ \bibinfo {year} {2007})\BibitemShut {NoStop}%
\bibitem [{\citenamefont {Efron}\ and\ \citenamefont
  {Tibshirani}(1994)}]{EfroTibs93}%
  \BibitemOpen
  \bibfield  {author} {\bibinfo {author} {\bibfnamefont {B.}~\bibnamefont
  {Efron}}\ and\ \bibinfo {author} {\bibfnamefont {R.}~\bibnamefont
  {Tibshirani}},\ }\href
  {https://www.crcpress.com/An-Introduction-to-the-Bootstrap/Efron-Tibshirani/p/book/9780412042317}
  {\emph {\bibinfo {title} {An Introduction to the Bootstrap}}},\ Chapman \&
  Hall/CRC Monographs on Statistics \& Applied Probability\ (\bibinfo
  {publisher} {CRC Press},\ \bibinfo {year} {1994})\BibitemShut
  {NoStop}%
\bibitem [{sup()}]{suppl}%
  \BibitemOpen
  \href@noop {} {\bibinfo {title} {Supplemental material},\
  }\bibinfo {note} {also on
  \href{http://www.indiana.edu/~jpac}{http://www.indiana.edu/\~{
  }jpac}}\BibitemShut {NoStop}%
\bibitem [{\citenamefont {Navarro~P\'erez}\ \emph {et~al.}(2015)\citenamefont
  {Navarro~P\'erez}, \citenamefont {Ruiz~Arriola},\ and\ \citenamefont {Ruiz~de
  Elvira}}]{Perez:2015pea}%
  \BibitemOpen
  \bibfield  {author} {\bibinfo {author} {\bibfnamefont {R.}~\bibnamefont
  {Navarro~P\'erez}}, \bibinfo {author} {\bibfnamefont {E.}~\bibnamefont
  {Ruiz~Arriola}}, \ and\ \bibinfo {author} {\bibfnamefont {J.}~\bibnamefont
  {Ruiz~de Elvira}},\ }\href {\doibase 10.1103/PhysRevD.91.074014} {\bibfield
  {journal} {\bibinfo  {journal} {Phys.Rev.}\ }\textbf {\bibinfo {volume}
  {D91}},\ \bibinfo {pages} {074014} (\bibinfo {year} {2015})},\ \Eprint
  {http://arxiv.org/abs/1502.03361}{\tt arXiv:1502.03361 [hep-ph]}\BibitemShut
  {NoStop}%
\bibitem [{\citenamefont {Tanabashi}\ \emph {et~al.}(2018)\citenamefont
  {Tanabashi} \emph {et~al.}}]{pdg}%
  \BibitemOpen
  \bibfield  {author} {\bibinfo {author} {\bibfnamefont {M.}~\bibnamefont
  {Tanabashi}} \emph {et~al.} (\bibinfo {collaboration} {Particle Data Group}
  Collaboration),\ }\href {\doibase 10.1103/PhysRevD.98.030001} {\bibfield
  {journal} {\bibinfo  {journal} {Phys.Rev.}\ }\textbf {\bibinfo {volume}
  {D98}},\ \bibinfo {pages} {030001} (\bibinfo {year} {2018})}\BibitemShut
  {NoStop}%
\end{thebibliography}%

\clearpage
\onecolumngrid
\section{Supplemental material}

\begin{table}[h] 
\caption{Parameters of the numerator $n^J_k(s)=\sum_{n=0}^3{a^{J,k}_n  T_n\left[\omega(s)\right]}$, with  
\mbox{$\omega(s)=s/(s+s_0)$},
 and $s_0= 1 \gev^2$ reflects the short range nature of $\etapi$ production. All numbers are expressed in\gev units. The first values are obtained from the best fit, and should be used to reproduce the plots. The second values contains the mean value and standard deviation estimated with $5\times 10^4$ bootstrapped datasets. We remark that the coefficients are $\gtrsim 95\%$ correlated, and the single error has to be taken with care. 
}
\begin{ruledtabular}
\begin{tabular}{c c c c c c}
\multicolumn{3}{c}{$\eta \pi$ channel}  & \multicolumn{3}{c}{$\eta' \pi$ channel} \\ \hline
$a^{P,\eta \pi}_0$ & $408.75$ & $356 \pm 334$ & $a^{P,\eta' \pi}_0$ & $-47.05$ & $-43 \pm 39$ \\
$a^{P,\eta \pi}_1$ & $-632.57$ & $-547 \pm 534$ & $a^{P,\eta' \pi}_1$ & $65.84$ & $59 \pm 63$ \\
$a^{P,\eta \pi}_2$ & $281.48$ & $240 \pm 255$ & $a^{P,\eta' \pi}_2$ & $-20.96$ & $-17 \pm 30$ \\
$a^{P,\eta \pi}_3$ & $-57.98$ & $-47 \pm 63$ & $a^{P,\eta' \pi}_3$ & $1.20$ & $-0 \pm 8$ \\
\hline
$a^{D,\eta \pi}_0$ & $-247.80$ & $-247 \pm 28$ & $a^{D,\eta' \pi}_0$ & $230.92$ & $233 \pm 79$ \\
$a^{D,\eta \pi}_1$ & $413.91$ & $415 \pm 39$ & $a^{D,\eta' \pi}_1$ & $-290.66$ & $-290 \pm 125$ \\
$a^{D,\eta \pi}_2$ & $-190.94$ & $-192 \pm 39$ & $a^{D,\eta' \pi}_2$ & $176.88$ & $177 \pm 83$ \\
$a^{D,\eta \pi}_3$ & $59.25$ & $61 \pm 29$ & $a^{D,\eta' \pi}_3$ & $-3.82$ & $-1 \pm 62$ \\
\end{tabular} 
\label{tab:num} 
\end{ruledtabular}
\end{table}

\begin{table}[h] 
\caption{Parameters of $D^J(s)$. The errors and correlations are estimated with bootstrap.}
\begin{ruledtabular}
\begin{tabular}{c c c c c c}
\multicolumn{3}{c}{Resonating terms}  & \multicolumn{3}{c}{$K$-matrix background} \\ \hline
$g^{P,1}_{\eta \pi}$ & $-0.68$ & $-0.55 \pm 0.38$ & $c^P_{\eta \pi,\eta \pi}$ & $-15.43$ & $-14.77 \pm 7.22$ \\
$g^{P,1}_{\eta' \pi}$ & $-13.12$ & $-13.12 \pm 0.95$ & $c^P_{\eta \pi,\eta' \pi}$ & $-67.22$ & $-65.28 \pm 13.91$ \\
$m^2_{P,1}$ & $3.52$ & $3.52 \pm 0.08$ & $c^P_{\eta' \pi,\eta' \pi}$ & $-190.73$ & $-184.19 \pm 38.21$ \\
 & & & $d^P_{\eta \pi,\eta \pi}$ & $1.82$ & $1.93 \pm 2.24$ \\
 & & & $d^P_{\eta \pi,\eta' \pi}$ & $7.64$ & $7.59 \pm 5.09$ \\
 & & & $d^P_{\eta' \pi,\eta' \pi}$ & $63.85$ & $60.54 \pm 18.59$ \\
\hline 
$g^{D,1}_{\eta \pi}$ & $5.63$ & $5.64 \pm 0.34$ & $c^D_{\eta \pi,\eta \pi}$ & $-2402.56$ & $-2385.05 \pm 273.87$ \\
$g^{D,1}_{\eta' \pi}$ & $-3.77$ & $-3.78 \pm 0.10$ & $c^D_{\eta \pi,\eta' \pi}$ & $462.60$ & $469.55 \pm 55.87$ \\
$m^2_{D,1}$ & $1.86$ & $1.86 \pm 0.02$ & $c^D_{\eta' \pi,\eta' \pi}$ & $-86.60$ & $-92.25 \pm 28.11$ \\
$g^{D,2}_{\eta \pi}$ & $147.79$ & $147.17 \pm 9.88$ & $d^D_{\eta \pi,\eta \pi}$ & $-614.58$ & $-608.35 \pm 49.32$ \\
$g^{D,2}_{\eta' \pi}$ & $-33.39$ & $-34.07 \pm 3.41$ & $d^D_{\eta \pi,\eta' \pi}$ & $164.72$ & $166.85 \pm 17.46$ \\
$m^2_{D,2}$ & $8.06$ & $8.06 \pm 0.30$ & $d^D_{\eta' \pi,\eta' \pi}$ & $-42.19$ & $-44.45 \pm 11.59$ \\
\end{tabular}
\label{tab:dmatrix} 
\end{ruledtabular}
\end{table}

\clearpage
\begin{table}[h] 
\caption{Summary of systematic studies of the denominator. For each systematic variation, $5\times 10^4$ bootstrapped pseudodatasets are produced, and the average is shown here. 
For each parameter varied, we consider the maximum deviation of the pole position from the one in the reference fit. If that is compatible with the statistical uncertainty, we neglect the effect. If larger, we assign a systematic uncertainty to it, and eventually add in quadrature all the systematic uncertainties. We vary the value of $s_L$ and $\alpha$ in the reference model in Eq. (2). As an alternative model, we use $\rho N^J_{ki}(s') = \delta_{ki}\, Q_J(z_{s'})\, s'^{-\alpha} \lambda^{-1/2} (s^{\prime},m^2_{\eta^{(\prime)}},m^2_\pi) $,
where $Q_J(z_{s'})$ is the second kind Legendre function, and $z_{s'}=1+2s' s_L/\lambda (s^{\prime},m^2_{\eta^{(\prime)}},m^2_\pi)$, with $s_L = 1\gev^2$.  
 Asymptotically it behaves  as $s'^{-\alpha}$, has a left hand cut starting at $s' = 0$, a short cut between $(s^{\prime}-m_{\eta^{(\prime)}})^2$ and $(s^{\prime}+m_{\eta^{(\prime)}})^2$, and an incomplete circular cut. 
}
\begin{ruledtabular}
 \begin{tabular}{c c | c c | c c}
Systematic & Poles & Mass \mevp & Deviation \mevp & Width \mevp & Deviation \mevp \\ \hline \hline
\multicolumn{6}{l}{~}\\ 
 \multicolumn{6}{c}{Variation of the function $\rho N(s')$ }\\ \hline\hline
\multirow{3}{*}{${s}_{{L}} = 0.8 \gev^{2}$} & $a_2(1320)$ & $1306.4$ & $0.4$ & $115.0$ & $0.6$ \\ 
 & $a_2'(1700)$ & $1720$ & $-3$ & $272$ & $26$ \\ 
 & $\pione$ & $1532$ & $-33$ & $484$ & $-8$ \\ 
\hline 
\multirow{3}{*}{${s}_{{L}} = 1.8 \gev^{2}$} & $a_2(1320)$ & $1305.6$ & $-0.4$ & $113.2$ & $-1.2$ \\ 
 & $a_2'(1700)$ & $1743$ & $21$ & $254$ & $7$ \\ 
 & $\pione$ & $1528$ & $-36$ & $410$ & $-82$ \\ 
\hline \multirow{3}{*}{Systematic assigned} & $a_2(1320)$ & & $0.0$ & & $0.0$ \\ 
 & $a_2'(1700)$ & & $21$ & & $26$ \\ 
 & $\pione$ & & $36$ & & $82$ \\ 
\hline 
\hline 
\multirow{3}{*}{$\alpha = 1$} & $a_2(1320)$ & $1305.9$ & $-0.1$ & $114.7$ & $0.3$ \\ 
 & $a_2'(1700)$ & $1685$ & $-37$ & $299$ & $52$ \\ 
 & $\pione$ & $1506$ & $-58$ & $552$ & $60$ \\ 
\hline \multirow{3}{*}{Systematic assigned} & $a_2(1320)$ & & $0.0$ & & $0.0$ \\ 
 & $a_2'(1700)$ & & $37$ & & $52$ \\ 
 & $\pione$ & & $58$ & & $60$ \\ 
\hline 
\hline 
\multirow{3}{*}{${Q}_{{J}}, \alpha = 1$} & $a_2(1320)$ & $1304.9$ & $-1.1$ & $114.2$ & $-0.2$ \\ 
 & $a_2'(1700)$ & $1670$ & $-52$ & $269$ & $22$ \\ 
 & $\pione$ & $1511$ & $-53$ & $528$ & $36$ \\ 
\hline 
\multirow{3}{*}{${Q}_{{J}}, \alpha = 1.5$} & $a_2(1320)$ & $1306.0$ & $0.1$ & $115.0$ & $0.6$ \\ 
 & $a_2'(1700)$ & $1717$ & $-5$ & $272$ & $25$ \\ 
 & $\pione$ & $1578$ & $14$ & $530$ & $39$ \\ 
\hline 
\multirow{3}{*}{${Q}_{{J}}, \alpha = 2$} & $a_2(1320)$ & $1306.2$ & $0.2$ & $114.7$ & $0.3$ \\ 
 & $a_2'(1700)$ & $1723$ & $1$ & $261$ & $15$ \\ 
 & $\pione$ & $1570$ & $6$ & $508$ & $16$ \\ 
\hline \multirow{3}{*}{Systematic assigned} & $a_2(1320)$ & & $1.1$ & & $0.0$ \\ 
 & $a_2'(1700)$ & & $52$ & & $25$ \\ 
 & $\pione$ & & $53$ & & $0$ \\ 
\hline 
\hline 
\multicolumn{6}{l}{~}\\ 
 \multicolumn{6}{c}{Variation of the numerator function $n(s)$ }\\ \hline\hline
\multirow{3}{*}{$\text{Polynomial expansion}$} & $a_2(1320)$ & $1305.9$ & $-0.1$ & $114.7$ & $0.3$ \\ 
 & $a_2'(1700)$ & $1723$ & $1$ & $249$ & $2$ \\ 
 & $\pione$ & $1563$ & $-1$ & $479$ & $-13$ \\ 
\hline \multirow{3}{*}{Systematic assigned} & $a_2(1320)$ & & $0.0$ & & $0.0$ \\ 
 & $a_2'(1700)$ & & $0$ & & $0$ \\ 
 & $\pione$ & & $0$ & & $0$ \\ 
\hline 
\hline 
\multirow{3}{*}{${t}_\text{eff} = -0.5 \gev^{2}$} & $a_2(1320)$ & $1306.8$ & $0.8$ & $114.1$ & $-0.3$ \\ 
 & $a_2'(1700)$ & $1730$ & $8$ & $259$ & $13$ \\ 
 & $\pione$ & $1546$ & $-18$ & $443$ & $-49$ \\ 
\hline \multirow{3}{*}{Systematic assigned} & $a_2(1320)$ & & $0.8$ & & $0.0$ \\ 
 & $a_2'(1700)$ & & $0$ & & $0$ \\ 
 & $\pione$ & & $0$ & & $0$ \\ 
\end{tabular}

\end{ruledtabular}
\end{table}

\begin{table}[h]
\caption{Summary of systematic studies of the numerator. For each systematic variation, $5\times 10^4$ bootstrapped pseudodatasets are produced, and the average is shown here. 
For each parameter varied, we consider the maximum deviation of the pole position from the one in the reference fit. If that is compatible with the statistical uncertainty, we neglect the effect. If larger, we assign a systematic uncertainty to it, and eventually add in quadrature all the systematic uncertainties. We vary the value of $t_\text{eff}$, and increase the order of the polynomial expansion by one unit.
}
\begin{ruledtabular}
\begin{tabular}{c c | c c | c c}
Systematic & Poles & Mass \mevp & Deviation \mevp & Width \mevp & Deviation \mevp \\ \hline \hline
\multicolumn{6}{c}{Variation of the numerator function $n(s)$ }\\ \hline\hline
\multirow{3}{*}{$\text{Polynomial expansion}$} & $a_2(1320)$ & $1305.9$ & $-0.1$ & $114.7$ & $0.3$ \\ 
 & $a_2'(1700)$ & $1723$ & $1$ & $249$ & $2$ \\ 
 & $\pione$ & $1563$ & $-1$ & $479$ & $-13$ \\ 
\hline \multirow{3}{*}{Systematic assigned} & $a_2(1320)$ & & $0.0$ & & $0.0$ \\ 
 & $a_2'(1700)$ & & $0$ & & $0$ \\ 
 & $\pione$ & & $0$ & & $0$ \\ 
\hline 
\hline 
\multirow{3}{*}{${t}_\text{eff} = -0.5 \gev^{2}$} & $a_2(1320)$ & $1306.8$ & $0.8$ & $114.1$ & $-0.3$ \\ 
 & $a_2'(1700)$ & $1730$ & $8$ & $259$ & $13$ \\ 
 & $\pione$ & $1546$ & $-18$ & $443$ & $-49$ \\ 
\hline \multirow{3}{*}{Systematic assigned} & $a_2(1320)$ & & $0.8$ & & $0.0$ \\ 
 & $a_2'(1700)$ & & $0$ & & $0$ \\ 
 & $\pione$ & & $0$ & & $0$ \\ 
\end{tabular}
\end{ruledtabular}
\end{table}

\end{document}